\documentclass[a4paper,fleqn,usenatbib]{mnras}

\usepackage{newtxtext}

\usepackage[T1]{fontenc}
\usepackage{ae,aecompl}

\usepackage{graphicx}	
\usepackage{amsmath}	
\usepackage{amssymb}	
\usepackage{multirow}
\usepackage{booktabs}
\usepackage{verbatim}
\usepackage{caption}
\usepackage{adjustbox}
\usepackage{hhline}
\usepackage{float,caption}
\usepackage{natbib}
\usepackage{titlesec}
\usepackage{subcaption}	
\usepackage[nameinlink,noabbrev]{cleveref}

\title[Improving Photometric Redshift Estimation]{Improving Photometric Redshift Estimation using {\sc GPz}: size information, post processing and improved photometry}

\author[Z. Gomes et al.]{Zahra Gomes,$^{1}$\thanks{E-mail: zahra.gomes@physics.ox.ac.uk}
Matt J. Jarvis,$^{1,2}$
Ibrahim A. Almosallam$^{3,4}$ 
and Stephen J. Roberts$^{3}$
\\
$^{1}$Oxford Astrophysics, Department of Physics, Keble Road, Oxford, OX1 3RH, UK\\
$^{2}$Department of Physics, University of the Western Cape, Bellville 7535, South Africa\\
$^{3}$Information Engineering, Parks Road, Oxford, OX1 3PJ, UK\\
$^{4}$King Abdulaziz City for Science and Technology, Riyadh 11442, Saudi Arabia\\
}

\date{Accepted XXX. Received YYY; in original form ZZZ}

\pubyear{2017}

\begin{document}
\label{firstpage}
\pagerange{\pageref{firstpage}--\pageref{lastpage}}
\maketitle

\begin{abstract}
The next generation of large scale imaging surveys (such as those conducted with the Large Synoptic Survey Telescope and {\em Euclid}) will require accurate photometric redshifts in order to optimally extract cosmological information. Gaussian Processes for photometric redshift estimation ({\sc GPz}) is a promising new method that has been proven to provide efficient, accurate photometric redshift estimations with reliable variance predictions. In this paper, we investigate a number of methods for improving the photometric redshift estimations obtained using {\sc GPz} (but which are also applicable to others). We use spectroscopy from the Galaxy and Mass Assembly Data Release 2 with a limiting magnitude of $r<19.4$ along with corresponding Sloan Digital Sky Survey visible ($ugriz$) photometry and the UKIRT Infrared Deep Sky Survey Large Area Survey near-IR (YJHK) photometry. We evaluate the effects of adding near-IR magnitudes and angular size as features for the training, validation and testing of {\sc GPz} and find that these improve the accuracy of the results by $\sim 15-20$ per cent. In addition, we explore a post-processing method of shifting the probability distributions of the estimated redshifts based on their Quantile-Quantile plots and find that it improves the bias by $\sim 40$ per cent. Finally, we investigate the effects of using more precise photometry obtained from the Hyper Suprime-Cam Subaru Strategic Program Data Release 1 and find that it produces significant improvements in accuracy, similar to the effect of including additional features. 
\end{abstract}

\begin{keywords}
methods: data analysis -- galaxies: distances and redshifts -- photometry  
\end{keywords}

\section{Introduction}

Large, deep redshift surveys are necessary for studying the large scale structure of the universe and the evolution of dark energy (\citealt{seo2003probing,hong2012correlation}), and a number of surveys have focused on achieving this goal [the Baryon Oscillation Spectroscopic Survey (BOSS) of the Sloan Digital Sky Survey III (SDSS-III; \citealt{dawson2012baryon}), the WiggleZ Dark Energy Survey (\citealt{blake2011wigglez2}), the 2df Galaxy Redshift Survey (\citealt{colless20012df}), the Kilo Degree Survey (KIDS, \citealt{de2013kilo}) and the Dark Energy Survey (DES; \citealt{dark2005dark})] and upcoming surveys such as {\em Euclid} (\citealt{Laureijs2011euclid}) and LSST (\citealt{lsstbook}) will provide unprecedented constraints on cosmological parameters. Although spectroscopic redshifts (hereafter spec-z, which we also denote as $z$) provide the most accurate redshift estimates, the process of obtaining spectroscopy is very time consuming, and is only feasible for nearby or bright galaxies, or very small areas containing faint galaxies (e.g. \citealt{alam2016clustering, lilly2009zcosmos}). Photometric redshifts (hereafter photo-z's, which we also denote as $\hat z$) on the other hand, provide a more efficient method of obtaining redshifts to much greater depths than possible for spectroscopy (\citealt{Connolly:aa, koo1985optical,blake2007cosmological,oyaizu2008galaxy}).

Therefore, cosmological measurements that use large redshift samples will benefit from the use of accurate photo-z's. One such cosmological measurement is the power spectrum of galaxies (or its Fourier transform: the two-point correlation function) which describes the distribution of galaxies on a range of scales(\citealt{hong2012correlation, alam2016clustering, jeong2015redshift, cole20052df,feldman1993power}), it is used for studying galaxy clustering, and on large scales, allows the detection of the Baryon Acoustic Oscillation feature\textemdash which provides measurements of the angular diameter distance and the Hubble parameter, thus placing constraints on the distance-redshift relation and the behaviour of dark energy (e.g. \citealt{Vargas-Magana:2016imr, alam2016clustering, hong2012correlation, lin2009observational}). Current and future large scale photometric surveys such as DES (\citealt{dark2005dark}), the Hyper Suprime-Cam Subaru Strategic Program (HSC SSP, \citealt{aihara2017first}) require accurate photo-z estimation methods to extract such cosmological information (\citealt{sanchez2014photometric,Tanaka2017}). Mixed photometric and spectroscopic surveys such as SDSS also benefit from photo-z estimation as photometry is always deeper than spectroscopy and allows the most efficient use of the survey data (e.g. \citealt{almosallam2016sparse, abdalla2011comparison, oyaizu2008galaxy, li2007estimating, blake2007cosmological}). Weak lensing studies also require large redshift samples and will also benefit from the larger samples that could be provided by accurate photo-z estimation methods (e.g. \citealt{hong2012correlation,jain2003cross,bridle2007dark,fu2008very,bernstein2010catastrophic}). As a result, a significant amount of work is being done to increase the efficiency and accuracy of the process via the creation of new algorithms and optimization of existing ones (e.g. \citealt{hildebrandt2010phat, abdalla2011comparison, benitez2009optimal, brammer2008eazy, hogan2015gaz, almosallam2016sparse}).

\subsection{Photo-z Estimation: Template Fitting Methods}
The method of using photometry to determine the redshift of galaxies was first developed in the 1960's by \citet{baum1962photoelectric}. This method involved using broad optical filters to collect the radiation from a galaxy followed by producing spectral energy distributions (SEDs). These SEDs were then compared to redshifted templates of the same galaxy type in the rest frame \textemdash using the transmission curve of the filters\textemdash to find the best fit and the corresponding redshift. Modern SED template fitting requires a library of either observed or synthetic SED templates of galaxies with stellar populations of various ages and for different star-formation histories. The observed fluxes are fitted to a linear combination of these templates, usually using a $\chi^2$ minimization procedure, to find the set of templates that provide the closest match and the corresponding redshift. The set of templates is chosen based on a number of factors such as star formation rate (SFR), metallicity, initial mass function (IMF), interstellar reddening, flux decreases due to the Lyman alpha forest and the limiting magnitude of each filter (e.g. \citealt{bolzonella2000photometric}). This method works because SEDs can be distinguished based on the shape of the continuum as well as the presence and location of strong spectral properties such as the 4000\AA\, break and strong emission lines [in the case of active galactic nuclei (AGN) and star forming galaxies (\citealt{bolzonella2000photometric})]. Some commonly used examples of template-fitting codes are {\sc Hyperz} (\citealt{bolzonella2000photometric}), {\sc Le Phare} (\citealt{ilbert2006accurate}) and {\sc EAZY} (\citealt{brammer2008eazy}).

The advantages of template fitting methods are that they allow easy extrapolation\textemdash allowing them to be used on very faint galaxies for which limited spectroscopy is available\textemdash and they also allow the determination of other physical properties of the galaxies, such as stellar mass and star-formation rate \citep[e.g.][]{Ilbert2015,Johnston2015}. However, a major drawback is the possibility of template mismatch due to template set incompleteness, this is particularly important considering that the templates are normally based on local galaxies, and thus do not necessarily represent galaxies in the entire sample \citep[e.g.][]{budavari2000creating, abdalla2011comparison}. Despite this, a library with too many galaxy templates can also be disadvantageous as it can result in colour-redshift degeneracies (\citealt{benitez2000bayesian}). SED template fitting is sometimes combined with Bayesian techniques such that galaxies with known spec-z's and similar properties to the galaxies being observed are used as priors to calibrate the templates (e.g. \citealt{ilbert2006accurate}). These methods often lead to improved results and also provide a probability density function that encompasses the uncertainty in the photo-z estimates. Examples of such methods are: {\sc ZEBRA} (\citealt{feldmannzurich}) and {\sc BPZ} (\citealt{benitez2000bayesian}). 

\subsection{Photo-z Estimation: Empirical and Machine Learning Methods}

Empirical techniques for photo-z estimation were first developed in the 1990's (e.g. \citealt{Connolly:aa,wang1998catalog}) and involved using a sample of galaxies with spectroscopic redshifts and photometric data to develop an empirical relationship between magnitude and redshift for a particular passband. In recent years, machine learning methods which develop complex models that fit the given data\textemdash making them superior to traditional empirical methods that are limited to simpler functions\textemdash have been developed (some examples are: {\sc ANNz} (\citealt{collister2004annz}), {\sc GAz} (\citealt{hogan2015gaz}), {\sc TPZ} (\citealt{kind2013tpz}) and {\sc GPz} (\citealt{almosallam2016sparse,Almosallam:2016aa}) which use artificial neural networks, genetic algorithms, random forests and Gaussian Processes, respectively). Machine learning methods require two independent datasets that contain both photometric data (and any other relevant data) and spectroscopic data, these are the training and validation sets.  The training set is used to develop the model by finding model parameters. As training is taking place, fitted models are run on the validation set to optimize the relevant parameters/weights and prevent overfitting to the training set. Finally, the resulting model is used for predicting the redshifts for a different set of galaxies given only their photometric data. In order to evaluate model performance, a third dataset with both photometric and spectroscopic data called the test set can be used, the model runs on the photometric data and the outputted results are compared to the known spectroscopic data.   

While the accuracy of photo-z's varies significantly depending on the method and the specific algorithm used, as well as the size and representativeness of the training set available, in general, these methods produce accurate redshift results coupled with acceptable estimates of uncertainty when a representative training set is available. Unlike SED fitting, these methods also do not require the nature of the observed galaxy to be explicitly known or assumed (\citealt{Connolly:aa}). On the other hand, the necessity of a representative spectroscopic sample is a major drawback: in the redshift desert, the lack of spectroscopic data would make it less likely to derive suitable empirical relations (\citealt{bolzonella2000photometric}) and similarly, the limited depth of spectroscopic data results in non-representative spectroscopic samples at high redshifts (\citealt{bolzonella2000photometric}). Thus, these methods normally outperform template-fitting methods when a spectroscopic sample that is large and representative is used, but perform poorly in comparison when such a sample is not available (such as at very faint magnitudes) (\citealt{oyaizu2008galaxy,firth2003estimating}). Therefore, some combination of these methods depending on the science goal is likely to be the most accurate.

In this paper we provide a brief overview of the {\sc GPz} algorithm and its advantages for photo-z estimation (\Cref{sec:GPz}), followed by a discussion on a number of approaches for improving the results obtained from {\sc GPz}. These include using near-IR photometric filters and the angular size of galaxies as inputs for the training, validation and testing of the {\sc GPz} model (\Cref{sec:YJHKsize}). We also investigate the use of a post-processing method that adjusts the positions of the probability distributions of the photo-z's\textemdash to minimize the deviation of the distributions obtained from those representative of the spectroscopic sample\textemdash based on their quantile-quantile plots  (\Cref{sec:optpdf}) and finally, the effect of photometric data with increased precision is discussed in \Cref{sec:impphot}. We provide conclusions to our work in \Cref{sec:conclusions}

\section{Photometric Redshift Estimation using {\sc GPz}}
\label{sec:GPz}

Gaussian Process (GP) regression (\citealt{rasmussen2006gaussian}) is a non-linear, Bayesian, non-parametric method of modelling distributions over functions. GP regression for photo-z estimation involves assuming that the input, $\mathbfit{x}_{i} \in \mathbb{R}^{d}$ (the set of $d$ magnitudes for the $i$-th object and\textemdash in the case of {\sc GPz}\textemdash the associated magnitude uncertainties) and output $y_i$ (the corresponding spec-z's) distributions are related such that:
\begin{equation}
y_i \sim \mathcal{N}\left(f(\mathbfit{x}_{i}), \sigma^{2}\right),\label{eq-gp-likelihood}
\end{equation}
assuming the following prior probability distribution over the function
\begin{equation}
 f(\mathbfit{x})\sim \mathcal{N}\left(0, \mathbfss{K}\left(\mathbfss{X},\mathbfss{X}\right)\right),\label{eq-gp-prior}
\end{equation}
where $\mathbfss{X}=\left\{\mathbfit{x}_{i}\right\}_{i=1}^{n} \in \mathbb{R}^{n \times d}$ is the set of $n$ training samples and $\mathbfss{K}\left(\mathbfss{X},\mathbfss{X}\right)\in\mathbb{R}^{n\times n}$ is a covariance function such that the element in the $i$-th row and the $j$-th column is determined by a function of the pair of inputs $\mathbfit{x}_{i}$ and $\mathbfit{x}_{j}$. The covariance function is unbounded, i.e. it expands with the size of the training set, and it captures our prior knowledge that close-by inputs should be mapped to close-by outputs; e.g. the squared exponential kernel $\mathbfss{K}(\mathbfit{x}_i,\mathbfit{x}_j)=\exp\left(-\left\|\mathbfit{x}_i-\mathbfit{x}_j\right\|^{2}/\lambda\right)$ for $\lambda>0$. From the likelihood in \Cref{eq-gp-likelihood} and the prior in \Cref{eq-gp-prior}, one can obtain the predictive probability distribution, using Bayes' theorem, for an unseen test case $\mathbfit{x}_{*}$ to be distributed as follows:

\begin{equation}
p(\mathbfit{x}_{*}|\mathbfit{y},\mathbfss{X},\sigma^{2})=\mathcal{N}\left(\mu_{*},\sigma_{*}^{2}\right)
\end{equation}
where $\mathbfit{y}=\left\{y_{i}\right\}_{i=1}^{n}\in\mathbb{R}^{n}$ is the set of $n$ outputs. Training the model then involves maximizing the probability of obtaining the outputs $\mathbfit{y}$ given the inputs $\mathbfss{X}$, this is done by maximizing the marginal likelihood $p(\mathbfit{y}|\mathbfss{X},\sigma^{2})$ (using the training and validation sets) which allows the determination of the optimal hyper-parameters ($\lambda$ and $\sigma^{2}$).

The mean function is then given by,
\begin{equation}\label{eq-predictive-mean}
\mu_{*}=\left(\mathbfss{K}\left(\mathbfss{X},\mathbfss{X}\right)+\mathbfss{I}_{n}\sigma^{2}\right)^{-1}\mathbfss{K}\left(\mathbfss{X},\mathbfit{x}_{*}\right)
\end{equation}
and the total variance, comprised of both the noise and model variance,
\begin{equation}\label{eq-predictive-variance}
\sigma_{*}^{2}=\nu_{*}+\sigma^{2}.
\end{equation}

This process has a large computational cost, $O(n^3)$, but the sparse Gaussian Process introduced by \citet{almosallam2016sparse} alleviates this problem by decreasing the number of kernel functions used without significantly reducing the accuracy of the regression model. In order to accomplish this \citet{almosallam2016sparse} allow each kernel function to have its own hyper-parameters in order to account for variable densities and patterns over the sample space, and the locations of these functions are optimized to represent the data distribution. 

\citet{almosallam2016sparse} also introduce cost-sensitive learning (CSL) methods, which allow the user to vary the weights and error functions of different regions of parameter space depending on the science goals the method is being used to achieve. One type of weighting that is utilized in the {\sc GPz} code is the normalization of the data points, these weights are defined as: 
\begin{equation}
\omega_i = \left(\frac{1}{1+z_{i}}\right)^2,
\end{equation}
where $\omega_i$ is the weight or error cost for sample $i$ and $z_{i}$ is the spec-z for sample $i$, thus giving lower redshift objects greater weight than higher redshift ones. In this analysis, we use the normalizing weights and no weights cases, with the application of these weights  termed CSL method `Normalized' and CSL method `Normal' respectively. 

The {\sc GPz} algorithm was further modified to address the problem of heteroscedastic (non-uniform, input-dependent) uncertainties in photometric data. The predictive variance obtained from GP regression, \Cref{eq-predictive-variance}, consists of two components, the model variance $\nu_*$ and the noise variance $\sigma^{2}$. The model variance describes the confidence level for the model that is fit to the data, this decreases as the density of the data in the colour-redshift space of a given data point increases. On the other hand, the noise uncertainty describes the spread of the data points in a given region of colour-redshift space, it therefore depends on the factors such as precision of the data and number of relevant features used. Noise uncertainty is normally assumed to be white gaussian noise, but in this case, in order to account for heteroscedastic noise, \citet{Almosallam:2016aa} model this term as a function of the input $\sigma^{2}(\mathbfit{x}_{*})$\footnote{\citet{Almosallam:2016aa} in practice model the precision not the variance, i.e. $\beta(\mathbfit{x})=1/\sigma^{2}(\mathbfit{x})$, for numerical concerns but we use the variance notation here for simplicity.} with its own hyper-parameters. This noise variance and the predictive mean function are then both learned over the optimization process. 

This sparse Gaussian Process method used for estimating photo-z's was found to outperform other selected machine learning methods in terms of performance metrics, reliability of variance measurements and the length of time required for training (\citealt{almosallam2016sparse, Almosallam:2016aa}). The incorporation of CSL methods allows optimal weighting of sample space and the separation of the variance terms enables the selection of galaxy samples based on both data sparsity and photometric noise in order to provide the most appropriate photo-z sample for a given science goal.

\section{Additional features for learning }
\label{sec:YJHKsize}

In this section we investigate how adding commonly available additional features, beyond optical colour/magnitude information, may help in improving the accuracy of photometric redshifts using {\sc GPz}. Similar studies have been done by \citet{Tagliaferri2003} and  \citet{Singal2011} which look at the effect of the addition of features such as galaxy morphology and size on neural network photo-z estimation methods. In addition, a comprehensive study of feature importance for photo-z estimation which included 85 derived or measured parameters such as magnitudes, colours, radii, morphology and ellipticity was presented by \cite{Hoyle2015}.  

\subsection{Near-IR magnitudes}

Large photometric redshift surveys often use photometric systems with 4-5 broad bands in the optical range (e.g. SDSS; \citealt{fukugita1996sloan}, DES; \citealt{dark2005dark}, PanStarrs; \citealt{chambers2016pan} and HSC; \citealt{aihara2017first}). Photometric redshift determination depends on the detection of continuum features in the SEDs of galaxies, and thus, localizing these features is important; for this reason, the traditional broad band filter systems are not necessarily ideal for photo-z estimation (\citealt{benitez2009optimal, budavari2001optimal}). Study of the Jacobian matrix of fluxes as a function of the physical properties of galaxies has shown that a spectral feature is most noticeable when the feature is in the overlap of two filters (\citealt{budavari2001optimal}). One way of improving the probability of this occurrence is to use narrower, more numerous filters (\citealt{hickson1994multinarrowband}), but this would require many more exposures, making it unfeasible. \citet{budavari2001optimal} explored the possibility of using an additional broad filter formed by combining multiple intermediate width filters located within the original broad filters. This method aids in the location of continuum features within the broad bands and does not significantly increase the total exposure time required as only one additional filter is added. Another study conducted by \citet{benitez2009optimal} experimented with the number of filters used, the degree of overlap among these filters and constant versus logarithmically increasing filter width. Some conclusions of the study were that for small numbers of filters the colour-redshift degeneracy prevents accurate estimations, particularly for faint galaxies. However, including near-IR data improved the photometric redshift accuracy as it reduced colour-redshift degeneracies. The system that was found to produce the best redshift depth and precision contained nine filters, logarithmically increasing filter width and halfwidth overlaps.

In this study, we will compare the use of the five ugriz filters (\citealt{fukugita1996sloan}) that span the optical range to the use of an additional four near-IR YJHK filters in the training of the {\sc GPz} algorithm. As mentioned previously, near-IR photometry decreases the effect of the colour-redshift degeneracy as it provides data from an additional portion of the spectrum. Near-IR photometry will also aid in the determination of redshifts of galaxies in the `redshift desert' ($1.2<z<1.8$) where the Balmer break is redshifted into this part of the spectrum (\citealt{rudnick2001k,mobasher2004photometric}), but we do not explore this in the present study. 

 \begin{table}
\centering
\begin{tabular}{ll}
  \hline
  Metric & Equation\\
  \hline
  RMSE & $\sqrt{\frac{1}{n}\sum^n_{i=1}\left(\frac{z_{i}-\hat z_{i}}{1+z_{i}}\right)^2}$\\
  BIAS & $\frac{1}{n}\sum^n_{i=1}\frac{z_{i}-\hat z_{i}}{1+z_{i}}$\\
  MLL & $\frac{1}{n}\sum^n_{i=1}-\frac{1}{2\sigma^2_i}\left(z_{i} - \hat z_{i}\right)^2 - \frac{1}{2}\ln\left(\sigma_i^2\right) - \frac{1}{2}\ln\left(2\upi\right)$\\
  FR$_{0.15}$ & $\frac{100}{n}|{i:\left|\frac{z_{i}-\hat z_{i}}{1+z_{i}}\right|<0.15}|$\\ 
  FR$_{0.05}$ & $\frac{100}{n}|{i:\left|\frac{z_{i}-\hat z_{i}}{1+z_{i}}\right|<0.05}|$\\   
   \hline
\end{tabular}
\caption{Equations defining the metrics used. Symbols $z_{i}$ and $\hat z_{i}$ are the spectroscopic and estimated photometric redshifts for source $i$ and $\sigma_i^2$ is the predictive variance.}
\label{tab:metric}
\end{table}

\subsection{Angular size}
\label{sec:size}

Machine learning methods are expected to benefit from using additional features if they provide relevant additional information that aids in determining the relationship between input and output variables, thus providing better constraints on the resulting model. These additional features are not necessarily magnitudes/colours as machine learning methods can take input data of different forms. The classification of the galaxy morphology of SDSS objects for the Galaxy Zoo project provides one example of this: the inputs to machine learning algorithms were not limited to the de-reddened colours, but other features that were related to morphology such as axis ratio measurements and log likelihoods from de Vaucouleurs and exponential fits were also incorporated (\citealt{banerji2010galaxy, gauci2010machine}). By the same token, the inputs of machine learning algorithms for photo-z estimation are not limited to magnitudes or fluxes. \citet{Tagliaferri2003,Hoyle2015,way2011galaxy} provide examples of the improvements to photo-z accuracy made by including features such as morphology and size as input. \citet{Singal2011} on the other hand found no significant improvement in photo-z estimates when shape parameters were added. In this analysis we will investigate the effects of inputting the angular size of galaxies, a relatively simple measurement to make for most astronomical data sets. 

Angular diameter distance, the ratio of the physical size of a body to the angular size we observe is related to redshift in such a way that for $z<1$, it is positively correlated with redshift. In this experiment, we exploit this relationship by inputing the angular sizes of the major and minor axes (measured in the $r$ band) of the observed galaxies as features for training.

\subsection{Experiment and Dataset}

The main data set used in this analysis consists of the ugriz and YJHK photometry, angular semi-major and semi-minor axis measurements and spectroscopic redshifts from the GAMA DR2 database (\citealt{liske2015galaxy}). The GAMA survey is a spectroscopic survey of 238,000 objects split into five survey regions covering a total area of ~286 deg$^2$ with a limiting magnitude of $r<19.8$ mag obtained using the AAOmega spectrograph on the Anglo-Australian Telescope. The second data release of GAMA contains the spectroscopic data along with photometric data and other additional information obtained from SDSS (which provided the ugriz magnitudes) and the UKIRT Infrared Deep Sky Survey Large Area Survey (UKIDSS-LAS) (which provided the YJHK magnitudes) for 72225 objects in three of the survey regions with total area 144 deg$^2$ with $r<19$ in two regions and $r<19.4$ in the third (\citealt{liske2015galaxy}). A sample of 63937 galaxies was obtained after removing all object duplicates, all objects with missing relevant data and all objects with normalized redshift quality (NQ $\leq 3$). Next, the data set was randomly split into training, validation and testing sets with a ratio of 2:2:1 and these sets were maintained for all experiments performed. 

The Gaussian Process with variable covariances (GP-VC) method was used with 100 basis functions and the modelling of heteroscedastic noise was included (\citealt{Almosallam:2016aa}). Experiments were done with the CSL methods normal and normalized (defined in \Cref{sec:GPz}), and for each of these, three datasets were used:  the standard five ugriz magnitudes, the nine ugrizYJHK magnitudes, and the ugrizYJHK magnitudes together with angular size data. For each set of results, five metrics were evaluated: the normalized root mean squared error (RMSE), the normalized bias (BIAS) the mean log likelihood (MLL) and the fraction retained with outlier thresholds 0.15 (FR$_{0.15}$) and 0.05 (FR$_{0.05}$). These are defined in \Cref{tab:metric}.  We also note that the addition of the near-IR and angular size features did not significantly increase the training time necessary (this remained under 2 minutes).

\subsection{Results and Analysis}

\Cref{fig:Original_results} shows scatter plots of photometric redshift versus spectroscopic redshift resulting from running the {\sc GPz} algorithm on the GAMA data (with SDSS/UKIDSS LAS photometry) using both CSL methods and the three sets of inputs. \Cref{tab:summary_alldata} gives the corresponding performance measures and predictive variances. These values are calculated for $0\le z\le0.6$ because the small training set at higher redshifts render the results unreliable. The straight line shown in the figures is the line of $z=\hat z$ and thus represents perfect prediction. Consistent improvement is seen as the near-IR magnitudes and then size data are added for both the normal and normalized methods as the distribution becomes tighter and lines up more symmetrically along the straight line. \Cref{tab:summary_alldata} also shows that the normal and normalized methods result in very similar performance measures.

The noise variance term decreased consistently as the additional features were added for both methods, this is as expected as adding additional, relevant features decreases the spread of the data points in the multidimensional colour-redshift space (\citealt{Almosallam:2016aa}). We see that the model variance also generally decreases with additional filters and size data. The model variance depends on the confidence about the model, which improves with data density. As features are added, the dimensionality of the model increases, and the data becomes more sparse, but if this additional data improves the model then this can counteract the decrease in data density and model variance can decrease. The normalized method also had lower noise variance values than the normal method in all cases. The normalized CSL method causes the model to preferentially fit the lower redshift regions, producing a completely different fit to what is obtained from the normal method. If the spread of the data is smaller in the lower redshift range, and the higher redshift range is not as important, then the resulting noise variance of the entire model (at all redshifts) can be lower than it would be using the normal method, thus explaining this result. For real situations, in which no spectroscopic data is present, the variance terms may be the only method of determining the quality of the results obtained, thus this general decrease of the variance with additional features is important as it corresponds to improved performance measures.

\newcommand{\imga}{\includegraphics[width=0.6\columnwidth,height=125pt]{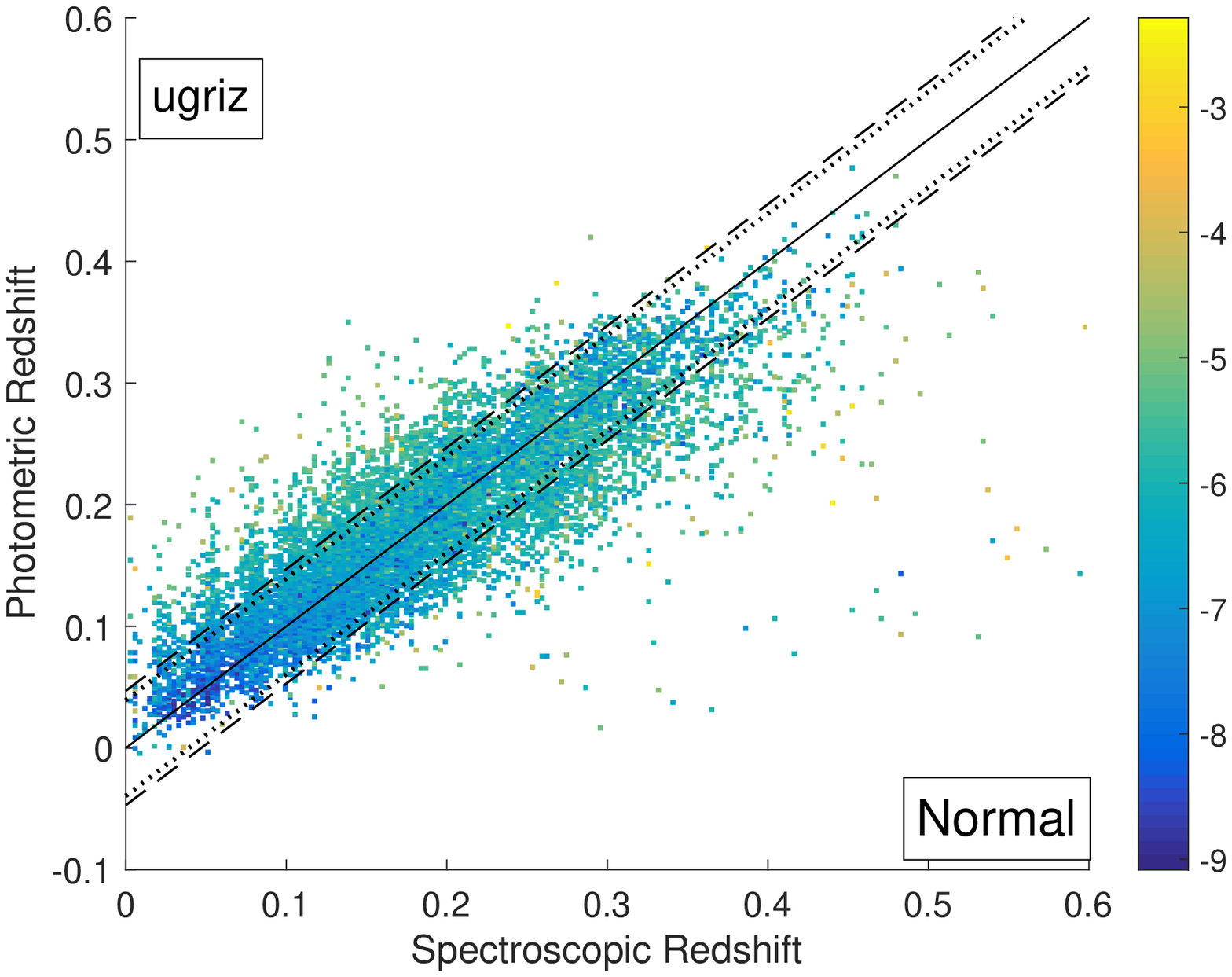}}
\newcommand{\imgb}{\includegraphics[width=0.6\columnwidth,height=125pt]{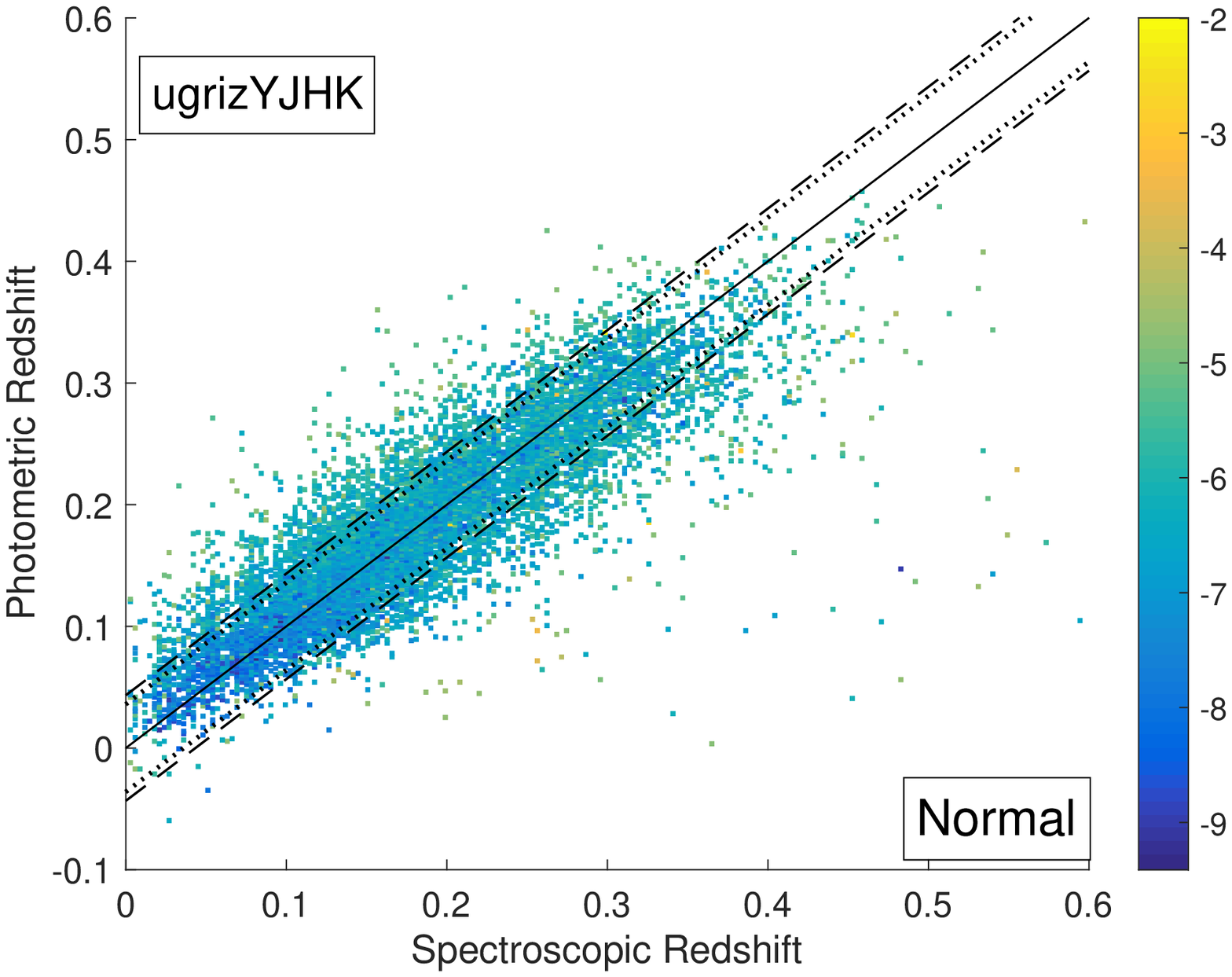}}
\newcommand{\imgc}{\includegraphics[width=0.6\columnwidth,height=125pt]{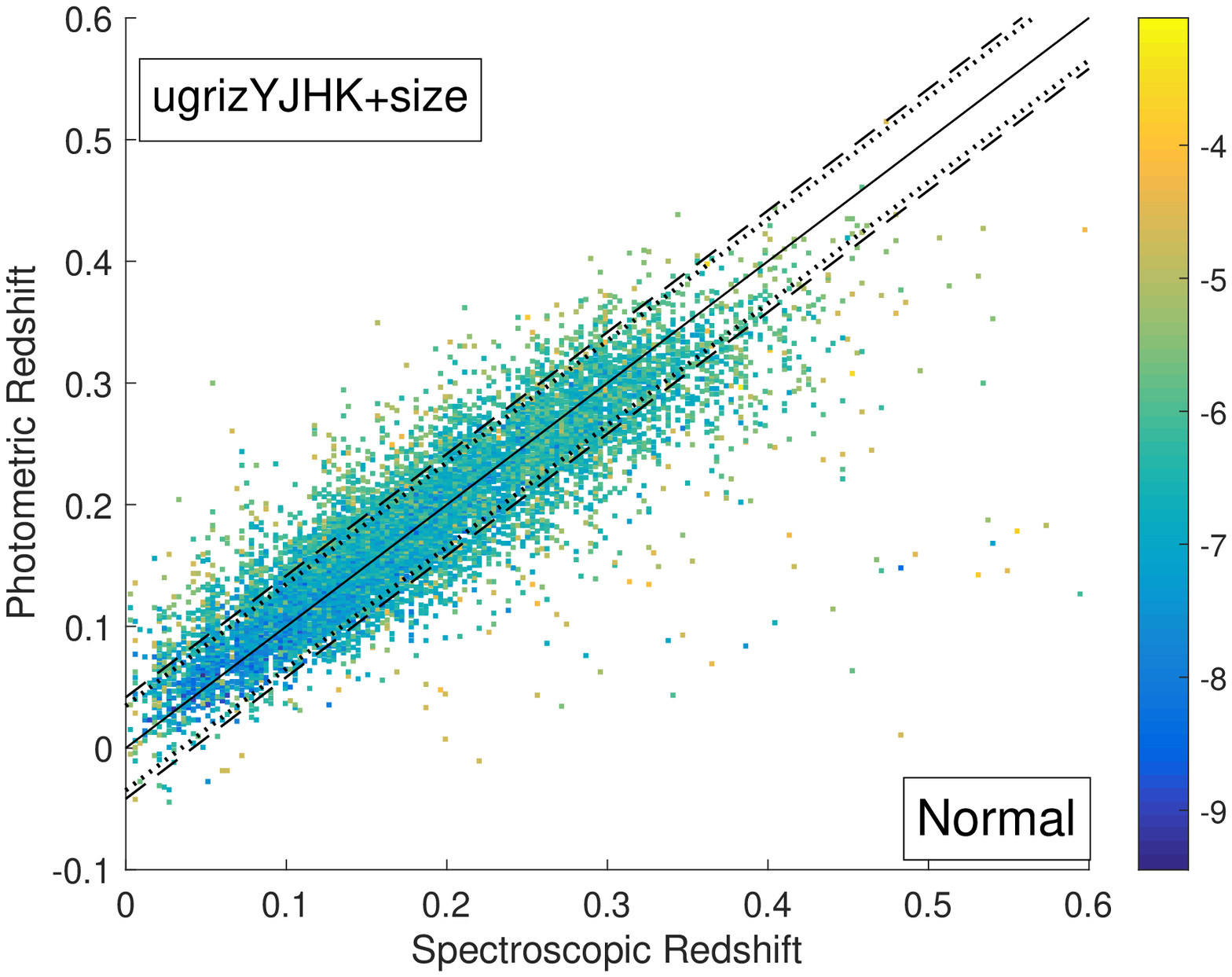}}
\newcommand{\imgd}{\includegraphics[width=0.6\columnwidth,height=125pt]{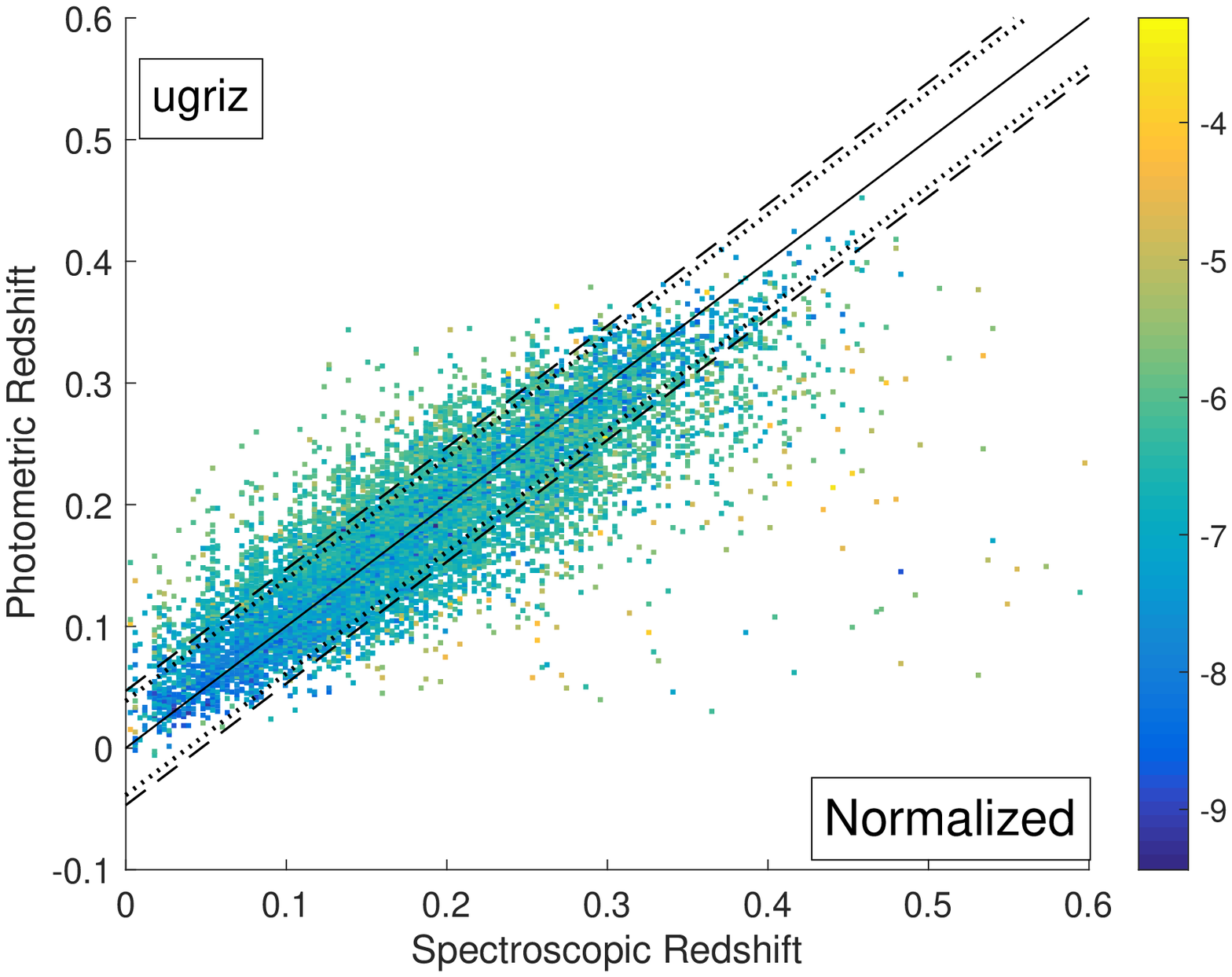}}
\newcommand{\imge}{\includegraphics[width=0.6\columnwidth,height=125pt]{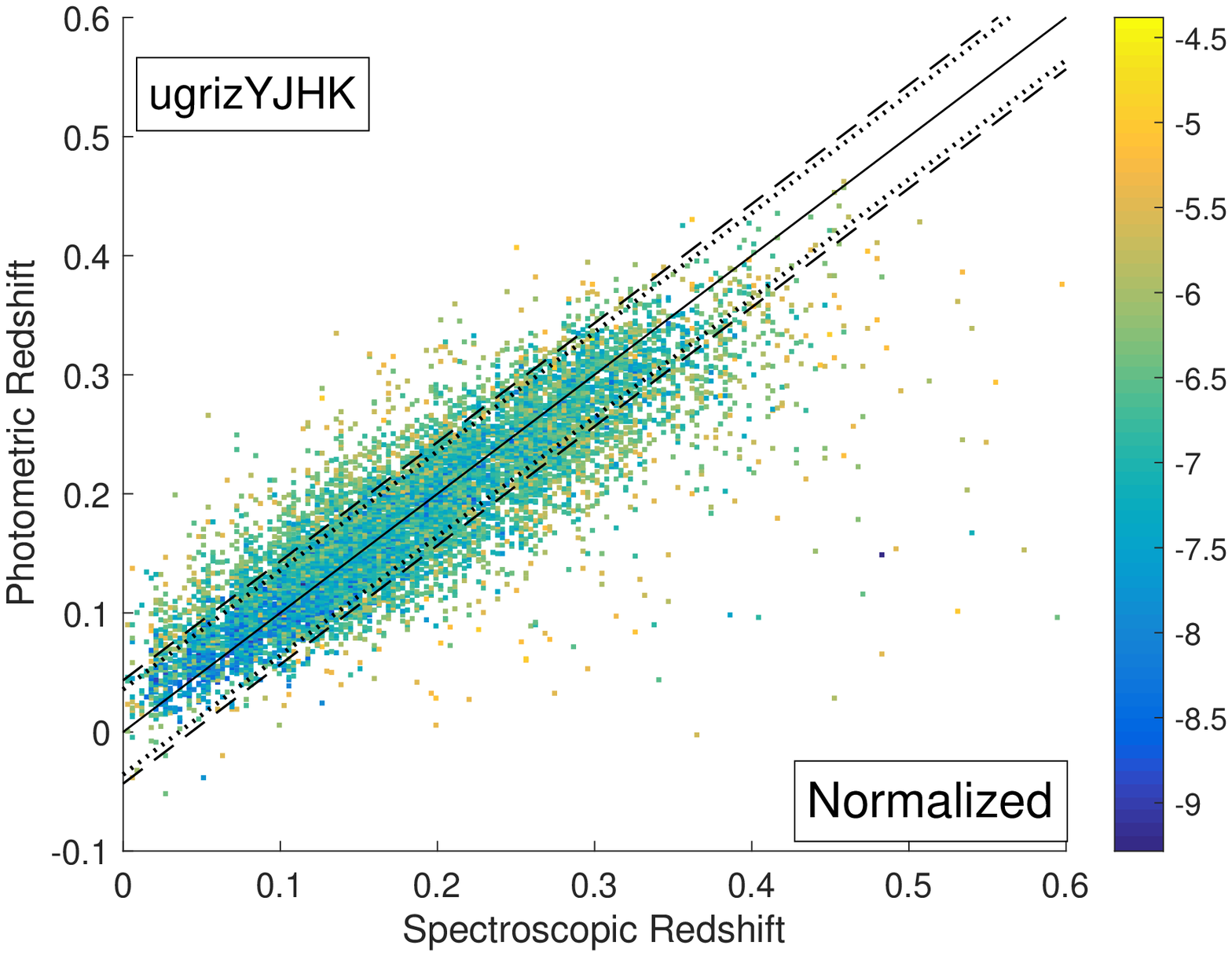}}
\newcommand{\imgf}{\includegraphics[width=0.6\columnwidth,height=125pt]{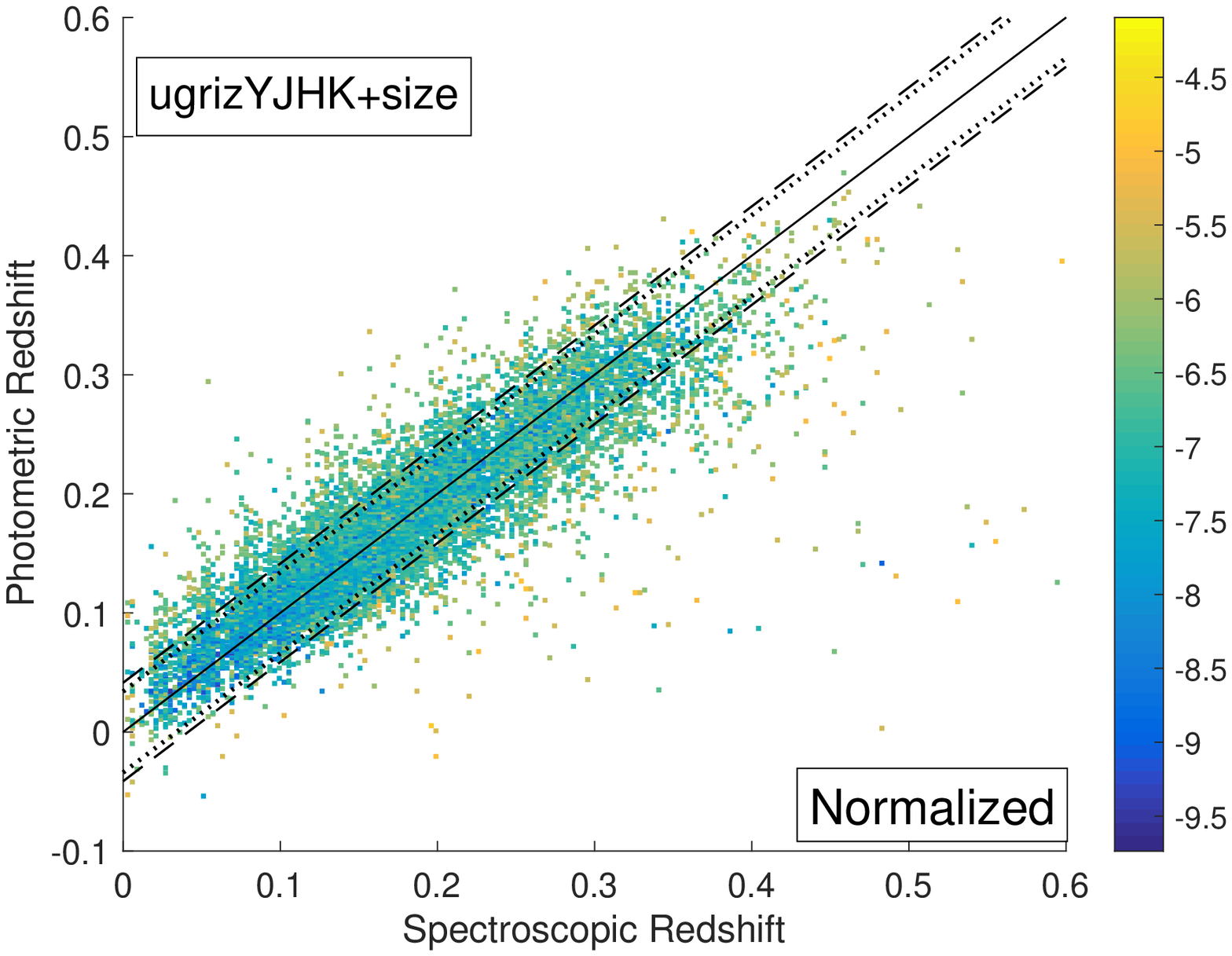}}

\begin{table*}
\centering
\begin{tabular}{ccc}

  \imga&\imgb&\imgc\\
  \imgd&\imge&\imgf\\

\end{tabular}
\captionof{figure}{Photometric redshift versus spectroscopic redshift plots using the CSL methods normal and normalized and using ugriz, ugrizYJHK and ugrizYJHK filters with size data. The colour scale represents the predictive variance, the solid line is the $z=\hat z$ line, and the dashed and dotted lines represent the rms scatter, $\sigma_{rms} =\sqrt{\frac{1}{n}\sum^n_{i=1}\left(z_{i}-\hat z_{i}\right)^2}$, and the normalized rms scatter, $\sigma_{rms} =\sqrt{\frac{1}{n}\sum^n_{i=1}\left(\left(z_{i}-\hat z_{i}\right)/\left(1+z_{i}\right)\right)^2}$, respectively, this is calculated using $0\le z\le0.6$.}
 \label{fig:Original_results}
\end{table*}

\newcommand{\imgaz}{\includegraphics[width=0.9\columnwidth]{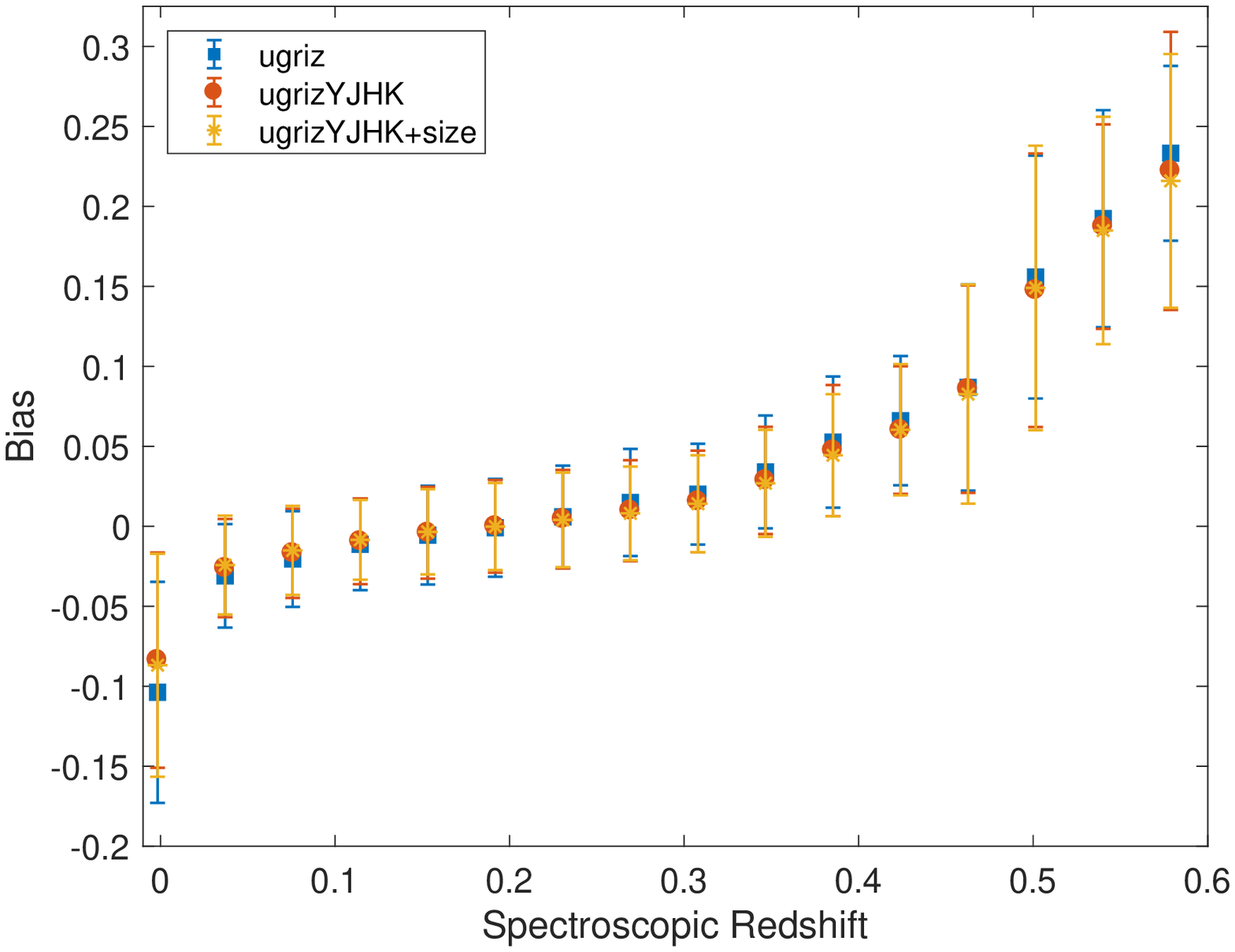}}
\newcommand{\imgbz}{\includegraphics[width=0.9\columnwidth]{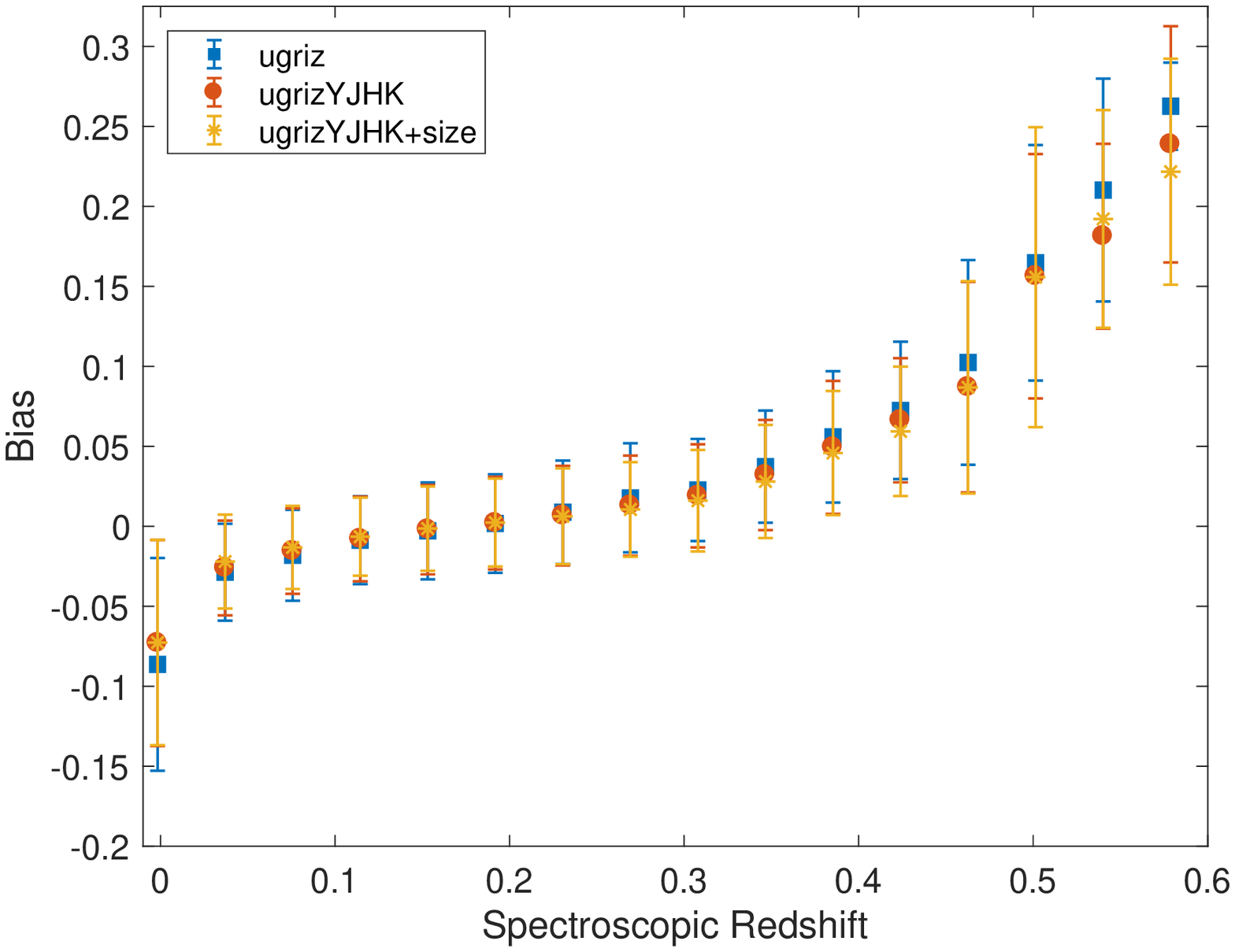}}

\begin{table*}
\centering
\begin{tabular}{cc}

  \imgaz&\imgbz

\end{tabular}
\captionof{figure}{ Bias versus redshift plots using the CSL methods normal (left) and normalized (right) and using ugriz, ugrizYJHK and ugrizYJHK filters with size data.}
 \label{fig:Bias}
\end{table*}

\begin{table*}
\centering
\begin{tabular}{llllllllll}
  \hline
 CSL Method & Filters & RMSE & BIAS & MLL & FR$_{0.15}$ & FR$_{0.05}$ & Variance & Model Var & Noise Var \\ 
  \hline
  \multirow{3}{*}{Normal}
 & ugriz          & 0.0393 & -0.00228 & 1.79 & 99.35  & 85.08  & 0.0023   & 7.0E-06   & 0.0023    \\
 & ugrizYJHK      & 0.0360 & -0.00185 & 1.84 & 99.54  & 87.92  & 0.0018   & 6.5E-06   & 0.0018    \\
 & ugrizYJHK+size & 0.0347 & -0.00177 & \textbf{1.91} & 99.50  & 89.61  & 0.0018   & 7.7E-06   & 0.0018    \\ 
  \hline
  \multirow{3}{*}{Normalized}  
& ugriz          & 0.0387 & 0.00069  & 1.73 & 99.38  & 85.09  & 0.0014   & 6.3E-06   & 0.0014    \\
& ugrizYJHK      & 0.0357 & \textbf{0.00031}  & 1.80 & 99.53  & 87.77  & 0.0013   & 6.9E-06   & 0.0013    \\
& ugrizYJHK+size & \textbf{0.0340} & 0.00048  & 1.85 & \textbf{99.55}  & \textbf{89.87}  & \textbf{0.0011}   & \textbf{6.2E-06}   & \textbf{0.0011}   \\
   \hline
   \end{tabular}
\caption{Summary performance measures and variances for the CSL methods normal and normalized with ugriz filters, ugrizYJHK filters and ugrizYJHK filters and size data. The number of training, validation and testing objects are: 25574, 25575 and 12788 respectively. The best metrics and variances are highlighted.}
\label{tab:summary_alldata}
\end{table*}

\begin{table*}
\centering

\begin{tabular}{llllllllllllll}
\toprule
\multicolumn{12}{c}{Normal} \\
\hline
               Redshift Bin & $\textnormal{N}_{\textnormal{train}}$ & $\textnormal{N}_{\textnormal{valid}}$ & $\textnormal{N}_{\textnormal{test}}$ & RMSE       & BIAS     & MLL         & $\textnormal{FR}_{0.15}$ & $\textnormal{FR}_{0.05}$  & Variance & Model Var & Noise Var  \\

               \hline
 0-0.1        & 4592   & 4621   & 2297  & 0.0510     & -0.0298 & 1.86   & 98.04  & 78.49  & 0.0020   & 9.2E-06        & 0.0020         \\
0.1-0.2      & 11613  & 11442  & 5888  & 0.0309     & -0.0063 & 2.02   & 99.92  & 89.93  & 0.0019   & 5.6E-06        & 0.0019         \\
0.2-0.3      & 6809   & 6973   & 3366  & 0.0345     & 0.0095  & 1.73   & 99.88  & 86.54  & 0.0028   & 7.1E-06        & 0.0028         \\
0.3-0.4      & 2117   & 2105   & 1012  & 0.0471     & 0.0313  & 1.25   & 99.31  & 74.41  & 0.0030   & 9.2E-06        & 0.0030         \\
0.4-0.5      & 259    & 244    & 134   & 0.0954     & 0.0766  & -2.20  & 91.04  & 37.31  & 0.0071   & 1.6E-05        & 0.0071         \\
0.5-0.6      & 27     & 28     & 14    & 0.2012     & 0.1879  & -10.22 & 35.71  & 0.00   & 0.0115   & 2.3E-05        & 0.0115         \\
\hline
\multicolumn{12}{c}{Normalized}     \\
             \hline
0-0.1        & 4592   & 4621   & 2297  & 0.0460     & -0.0263 & 1.86   & 98.65  & 81.15  & \textbf{0.0013}   & 7.5E-06        & 0.0013         \\
0.1-0.2      & 11613  & 11442  & 5888  & \textbf{0.0299}     & \textbf{-0.0036} & \textbf{2.05}   & \textbf{99.93}  & \textbf{90.64}  & \textbf{0.0013}   & \textbf{5.1E-06}        & \textbf{0.0012}         \\
0.2-0.3      & 6809   & 6973   & 3366  & 0.0357     & 0.0123  & 1.68   & 99.79  & 84.28  & 0.0017   & 6.4E-06        & 0.0017         \\
0.3-0.4      & 2117   & 2105   & 1012  & 0.0493     & 0.0341  & 0.89   & 98.91  & 72.43  & 0.0018   & 8.6E-06        & 0.0018         \\
0.4-0.5      & 259    & 244    & 134   & 0.1039     & 0.0858  & -4.89  & 87.31  & 33.58  & 0.0036   & 1.6E-05        & 0.0036         \\
0.5-0.6      & 27     & 28     & 14    & 0.2194     & 0.2067  & -16.24 & 35.71  & 0.00   & 0.0054   & 2.2E-05        & 0.0054        \\
               \bottomrule
\end{tabular}
\caption{Table showing performance measures and variances by redshift bin for the CSL methods normal and normalized with ugriz filters. The best metrics and variances are highlighted.}
\label{tab:alldata_normal}
\end{table*}

\newcommand{\imgza}{\includegraphics[width=0.9\columnwidth]{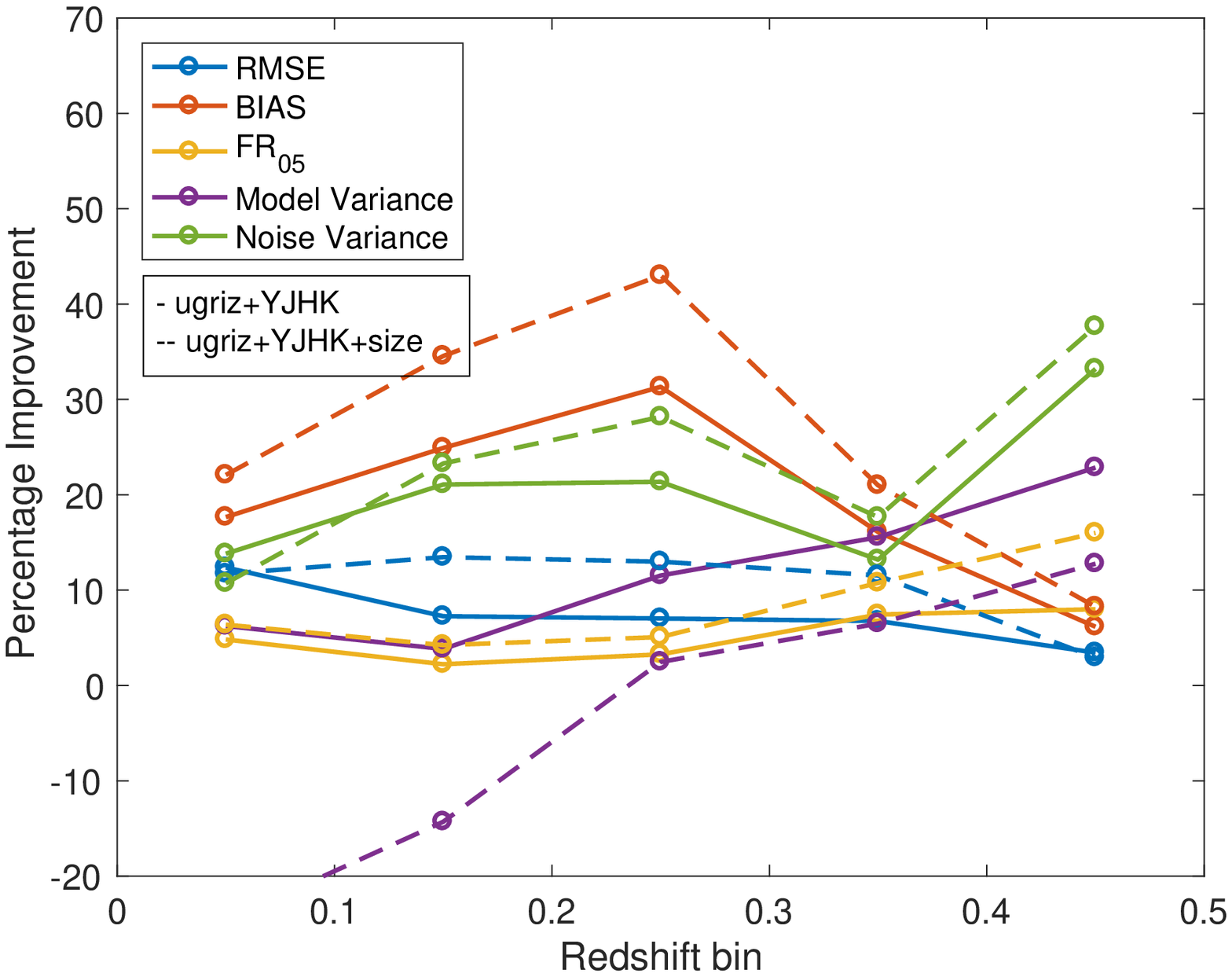}}
\newcommand{\imgzb}{\includegraphics[width=0.9\columnwidth]{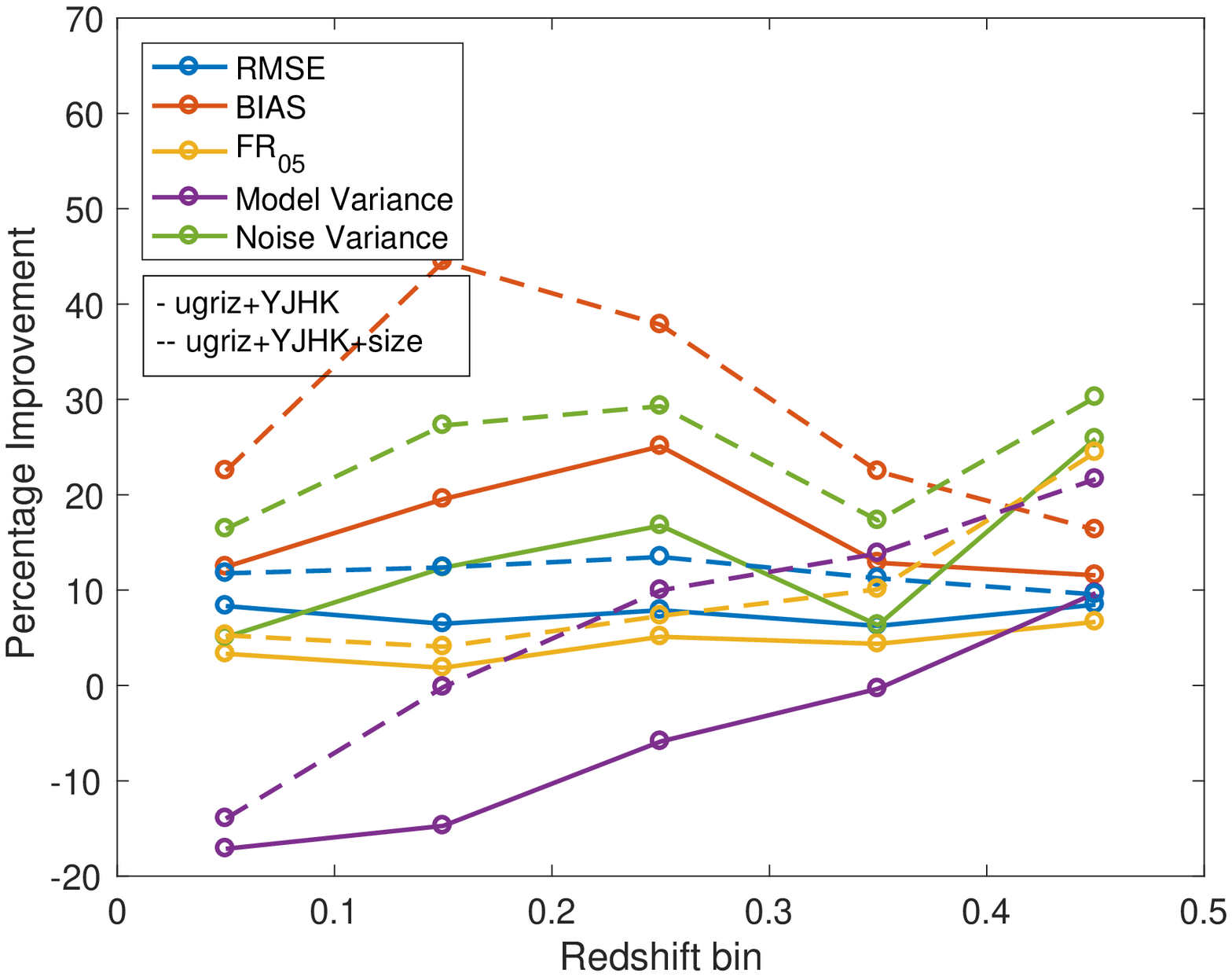}}

\begin{table*}
\centering
\begin{tabular}{cc}

  \imgza&\imgzb

\end{tabular}
\captionof{figure}{Percentage improvements of performance measures and variances by redshift bin due to the use of ugrizYJHK filters and size data using the normal (left) and normalized (right) methods. The solid lines represent the improvements due to adding the near-IR features and the dashed lines represent the improvements due to adding both the near-IR and angular size features.}
 \label{fig:percent_improvements}
\end{table*}

Next, we defined redshift bins of width 0.1 and calculated the five metrics and average variances for each redshift bin in order to understand the relationships between the CSL methods, features used and redshift range. \Cref{tab:alldata_normal} shows these metrics using ugriz features, and the same trends are observed when the additional features were added. It is clear that the results improve as the number of objects in the redshift bin increases: the 0.1-0.2 bin contained the largest number of objects and correspondingly produced the best results, the bin with fewest data points (0.5-0.6) produced the poorest results. This is because the {\sc GPz} code minimizes the total sum of squared errors, and therefore will preferentially fit the regions of sample space with higher densities of data points. The normalized method performed better than the normal method at lower redshifts ($0 <z<0.2$), while the normal method performed better in the higher redshift regions ($0.2 < z< 0.6$). This is expected since the normalized method gives more weight to the lower redshift objects than the higher redshift ones in the training of the model. This effect of the normalizing weights implies that this method would not be appropriate for science goals which require accurate photometric redshifts of higher redshift galaxies for which data is scarce. On the other hand, if a sample is expected to be mostly at low redshifts, and the accuracy of the few high redshift objects is not important then the normalized method will provide some added accuracy in the low redshift regime. When analysing the variance estimates by redshift bin we find that in all redshift bins, the normalized method again resulted in noise variance values that were lower than for the normal method. The model variance decreases with increasing number of objects in each redshift bin, this is expected behaviour for the model uncertainty as it decreases as the data density increases. On the other hand, the noise variance increases with increasing redshift, this is because although the higher redshift regions contain less training data, the photometry is likely to be less accurate as these galaxies are fainter on average, leading to a larger spread of the estimated photo-z's.

\Cref{fig:percent_improvements} shows the percentage improvements resulting from adding the near-IR followed by the angular size features. Improvements are clearly seen across all metrics in all redshift bins and apart from one case involving the model variance, the angular size features clearly provide significant added improvements compared to the near-IR features alone. The RMSE and $\textnormal{FR}_{0.05}$ metrics both undergo smaller improvements in regions with higher data densities, where the original estimates were more accurate, while they increase more significantly ($\textnormal{FR}_{0.05}$ in particular) in the lower density regions. The bias appears to undergo significant improvements over all redshift bins, but because the original values were very small (see \Cref{tab:alldata_normal}), very small changes led to large percentage improvements. \Cref{fig:Bias} shows the bias as a function of redshift, and here, a similar trend to the other metrics is observed: more significant improvements occur in regions of lower number densities. Improvements of $\textnormal{FR}_{0.15}$ (not shown) were negligible, while those of $\textnormal{FR}_{0.05}$ were more significant in all redshift bins, this implies that the addition of these features does not have a significant influence on the worst outliers, but decreases the scatter of objects with smaller initial deviations.

\subsection{Outliers}

\begin{figure}
\centering
\includegraphics[width=0.8\columnwidth]{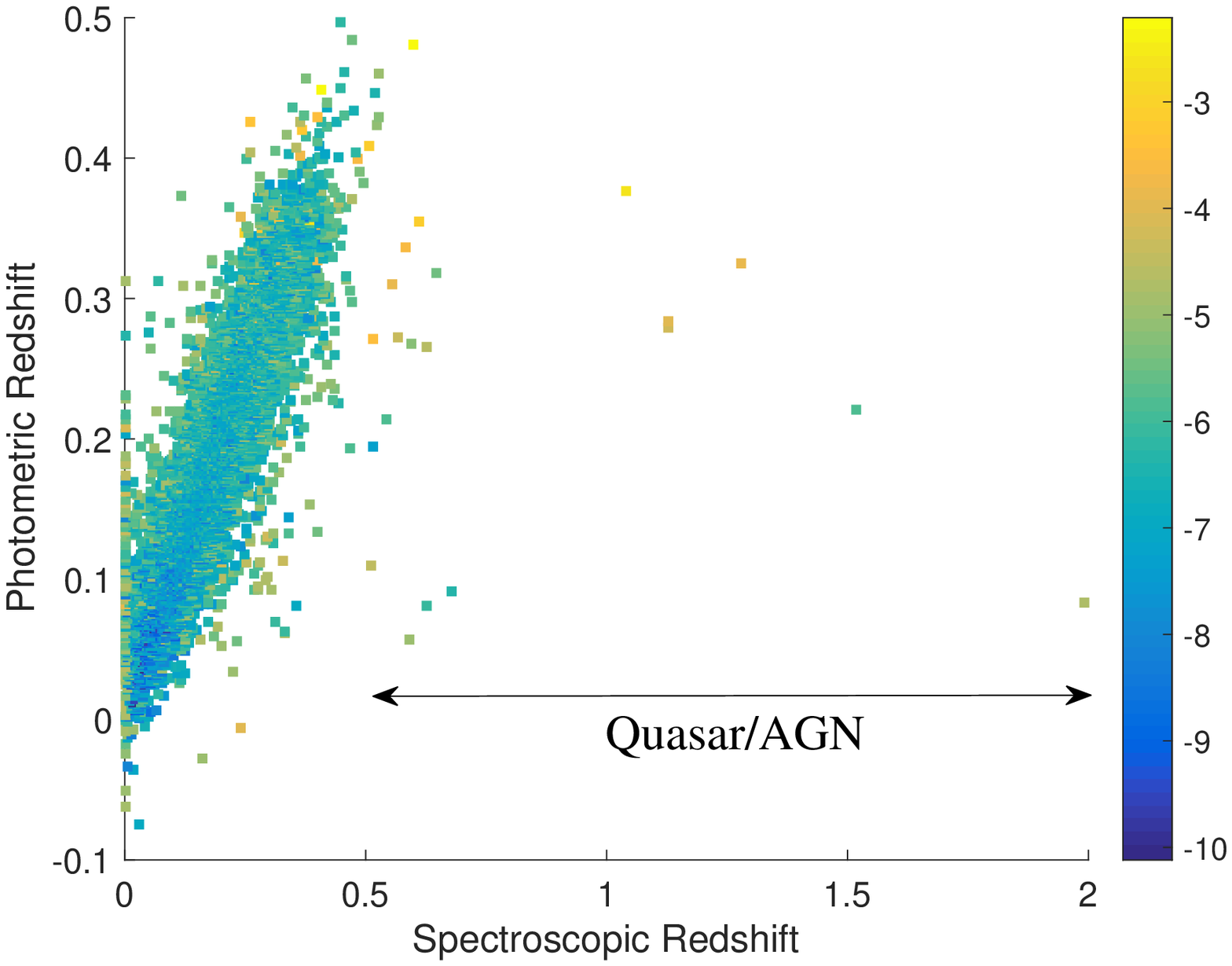}
 \caption{Plot showing the location of catastrophic outliers.}
 \label{fig:Quasars}
\end{figure}

\begin{figure}
\centering
\includegraphics[width=0.9\columnwidth]{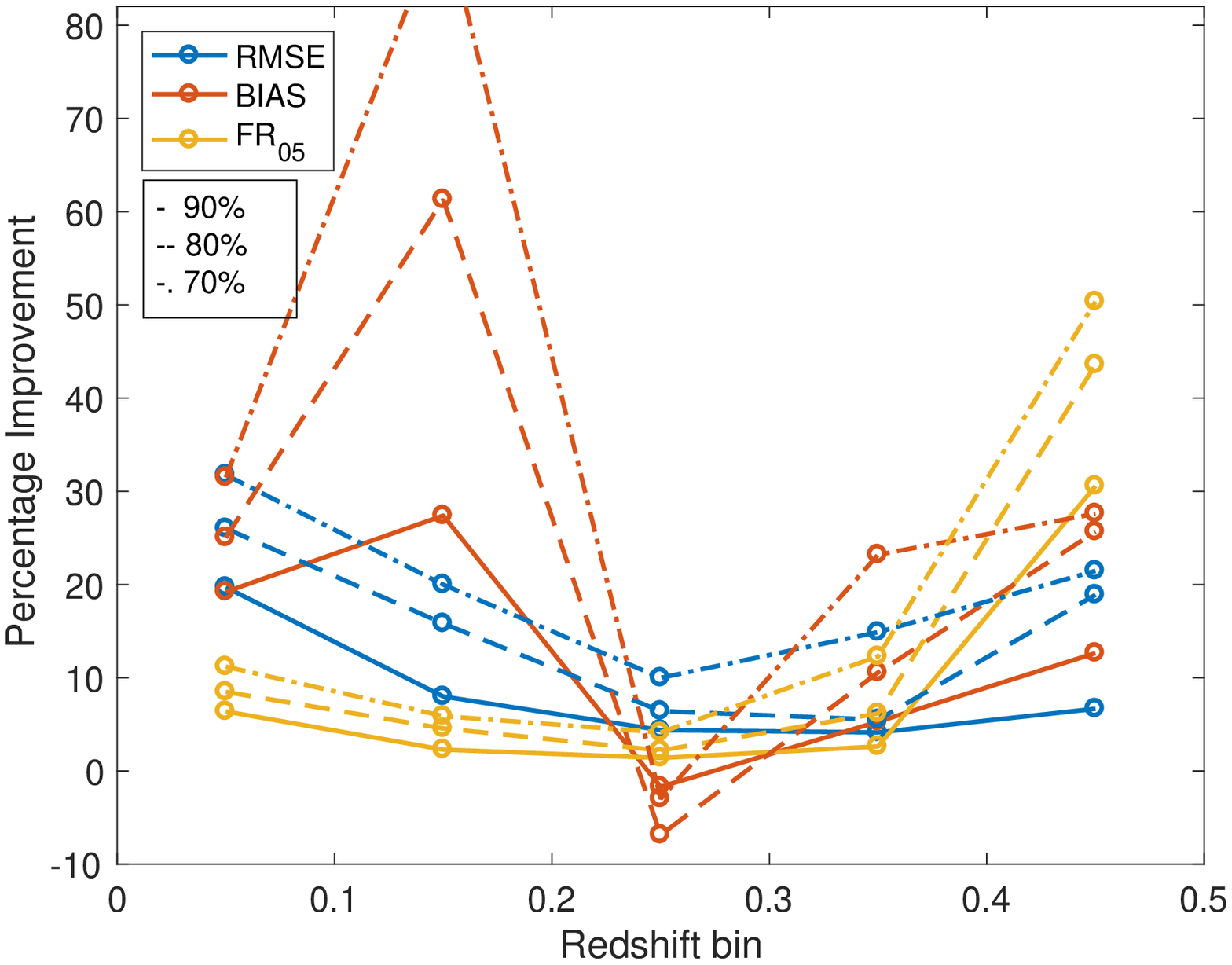}
 \caption{Percentage improvements of performance measures by redshift bin when 90, 80 and 70 per cent of the testing data with the lowest uncertainties were used. CSL method normal using ugrizYJHK filters and size data was used}
 \label{fig:sigma_cuts}
\end{figure}

When the entire redshift range for which data is present ($z<2.1$) was studied, we identified those objects with the highest fractional errors $\left(\frac{|z-\hat z|}{1+z}>0.15\right)$ and found that most of these objects were either quasars, narrow-line AGN or had noisy spectra that made it difficult to determine a redshift. In addition, although this set contained objects with a range of redshifts, all the outliers with high redshifts ($z>0.55$) were contained in this group, the positions of these outliers are shown in \Cref{fig:Quasars}.

The reason why the {\sc GPz} algorithm was unable to correctly predict the redshift of these quasars and AGN is because too few of these were present in the training data to allow the algorithm to make realistic estimates. We see from the analysis in the previous section that the number of objects available for training in each bin is an important factor in obtaining accurate estimates, thus if a large sample of quasars and other AGN was present, we expect that photo-z estimation of these objects would be greatly improved. For the objects with noisy spectra, the spectroscopic redshifts may have been incorrectly determined (as our constraint of NQ $\leq 3$ will not result in 100 per cent accuracy for the spectroscopic redshifts), in which case the photometric redshift estimate may be more accurate than the spectroscopic redshift. 

In \Cref{fig:sigma_cuts} we show the improvement in metric performance as we reduce the sample according to the variance prediction. We see that as higher variance estimates are removed, our results improve greatly: cutting the values with higher uncertainties and using 90, 80 and 70 per cent of the test data with the lowest variances shows consistent improvement in all metrics. This was using using ugrizYJHK filters and size data, but the same trend is observed using only ugriz and ugrizYJHK filters. As this removal of data was based solely on the variance values, this method can also be used for real surveys to obtain appropriate samples\textemdash based on the specifications for data density and variance necessary for the specific science case.

\section{Optimizing the Probability Density Functions}
\label{sec:optpdf}

It has become clear that single point estimates of the photometric redshifts are insufficient for many scientific applications, and the full probability density function (PDF) is preferred. However, it is extremely difficult to obtain reliable PDFs from both template fitting (due to non-representative templates) and empirical methods (where for example absence of data is traditionally difficult to quantify). Some methods employ post-processing to give their estimated PDFs the correct statistical properties (see \citealt{Bordoloi2012,Polsterer2016}), {\sc GPz} overcomes this by introducing an additional noise term to alleviate some of these issues.

\begin{figure*}
	
	\begin{subfigure}[b]{0.3\textwidth}
		\centering
                 \includegraphics[width=\textwidth]{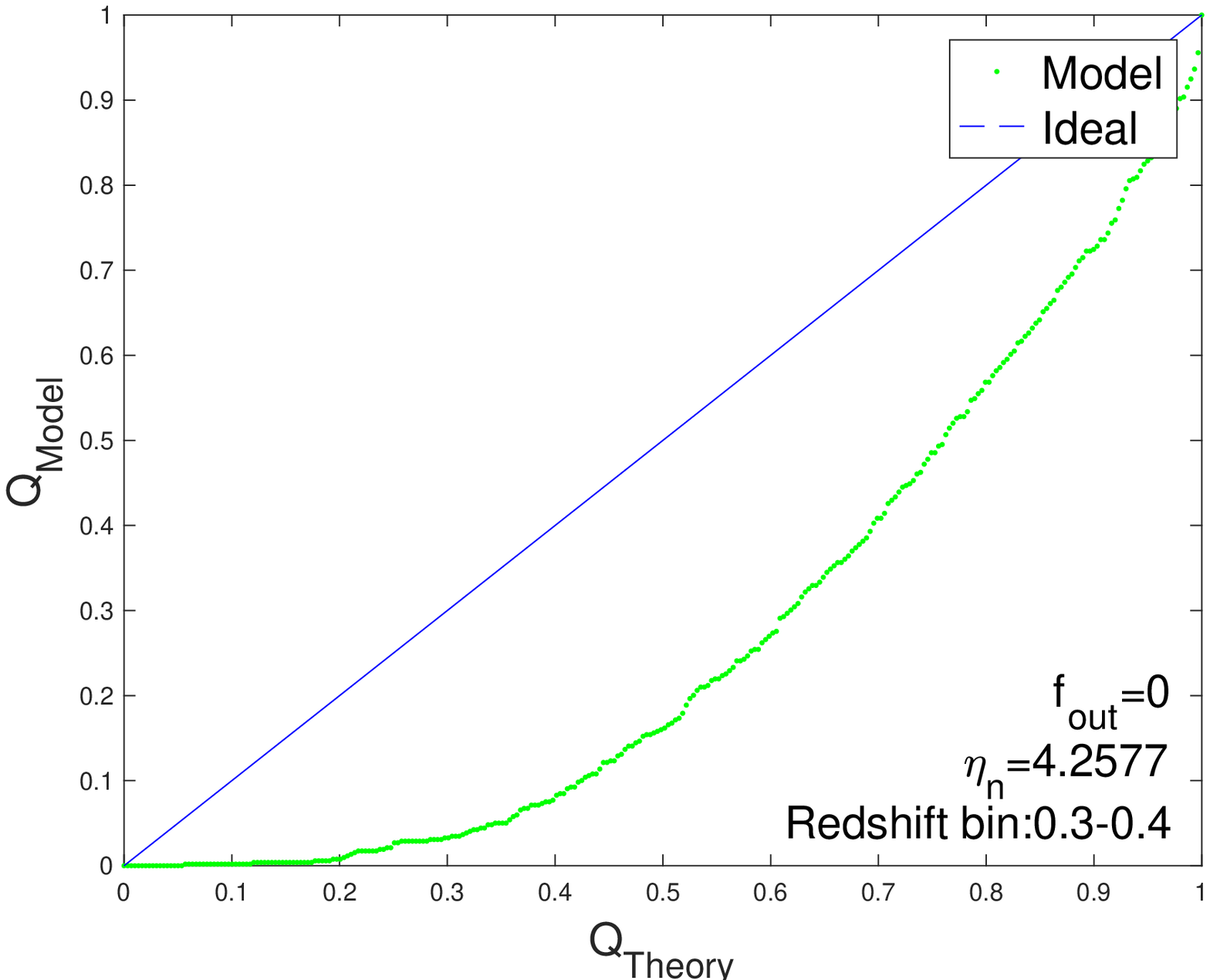}
                 \caption{Before shift}\label{subfig:Q-Q_before}
        \end{subfigure}
        ~
        \begin{subfigure}[b]{0.3\textwidth}
		\centering
                 \includegraphics[width=\textwidth]{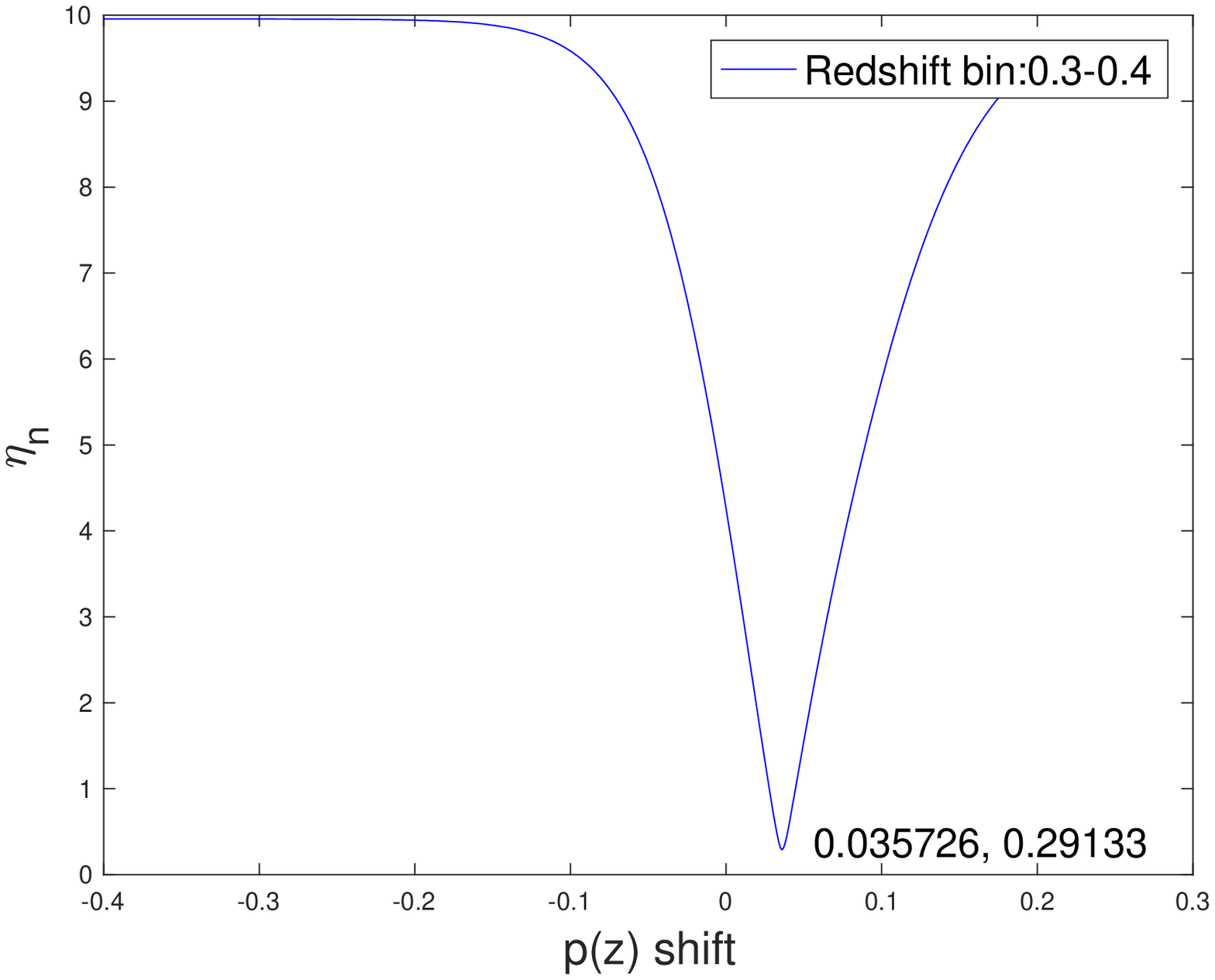}
                 \caption{$\eta_n$ vs $P(z)$}\label{subfig:eta-vs-Pz}
        \end{subfigure}
        ~
        \begin{subfigure}[b]{0.3\textwidth}
		\centering
                 \includegraphics[width=\textwidth]{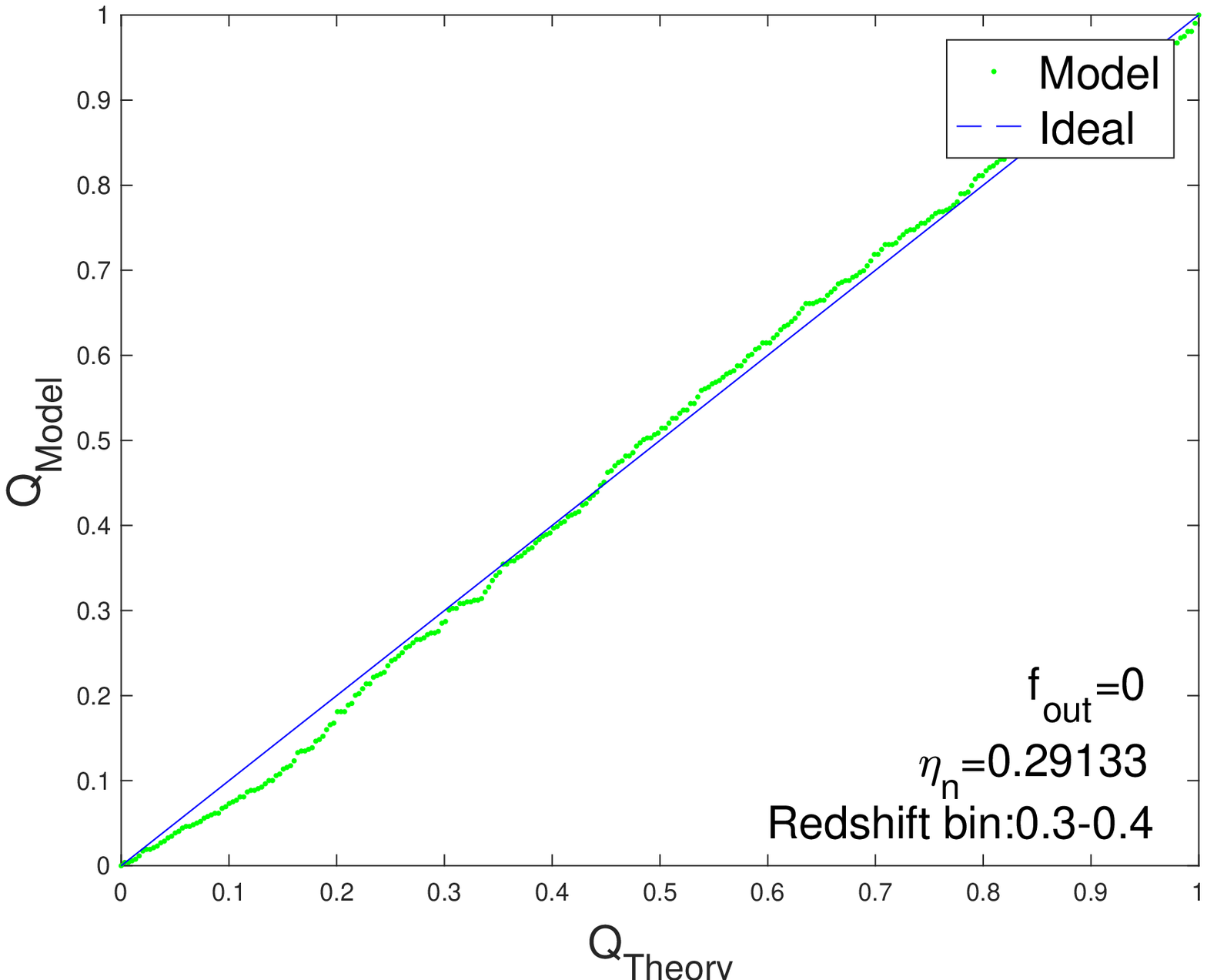}
                 \caption{After shift}\label{subfig:Q-Q_after}
        \end{subfigure}
        
        \caption{(a) Q-Q plot in the redshift bin 0.3-0.4 with the CSL method normal using ugriz filters before applying shifts, (b) the corresponding $\eta_n$ vs $P(z)$ and (c) the Q-Q plot after the $p(z)$ shift was applied.}
        \label{fig:badexample}
\end{figure*}

In this section we investigate the accuracy of the probability density functions (PDFs) of the photometric redshifts using Quantile-Quantile (Q-Q) plots (as in \citealt{Wittman2016}). We use these to provide appropriate alterations to the PDFs with the aim of further optimizing the redshift estimates. The first step in doing this is calculating the percentiles of the spectroscopic redshifts relative to the PDFs of the photometric redshifts. The PDF of the photo-z for a given object obtained using the {\sc GPz} algorithm is Gaussian with mean equal to the photo-z estimate and variance given by the sum of the model and noise variances. The percentile of a given redshift ($z_1$) is given by the value of the cumulative distribution function (CDF) at that redshift [CDF($z=z_1$)]. In this way, the percentiles of every spec-z relative to the photo-z PDFs were calculated.

\newcommand{\imgze}{\includegraphics[width=0.9\columnwidth]{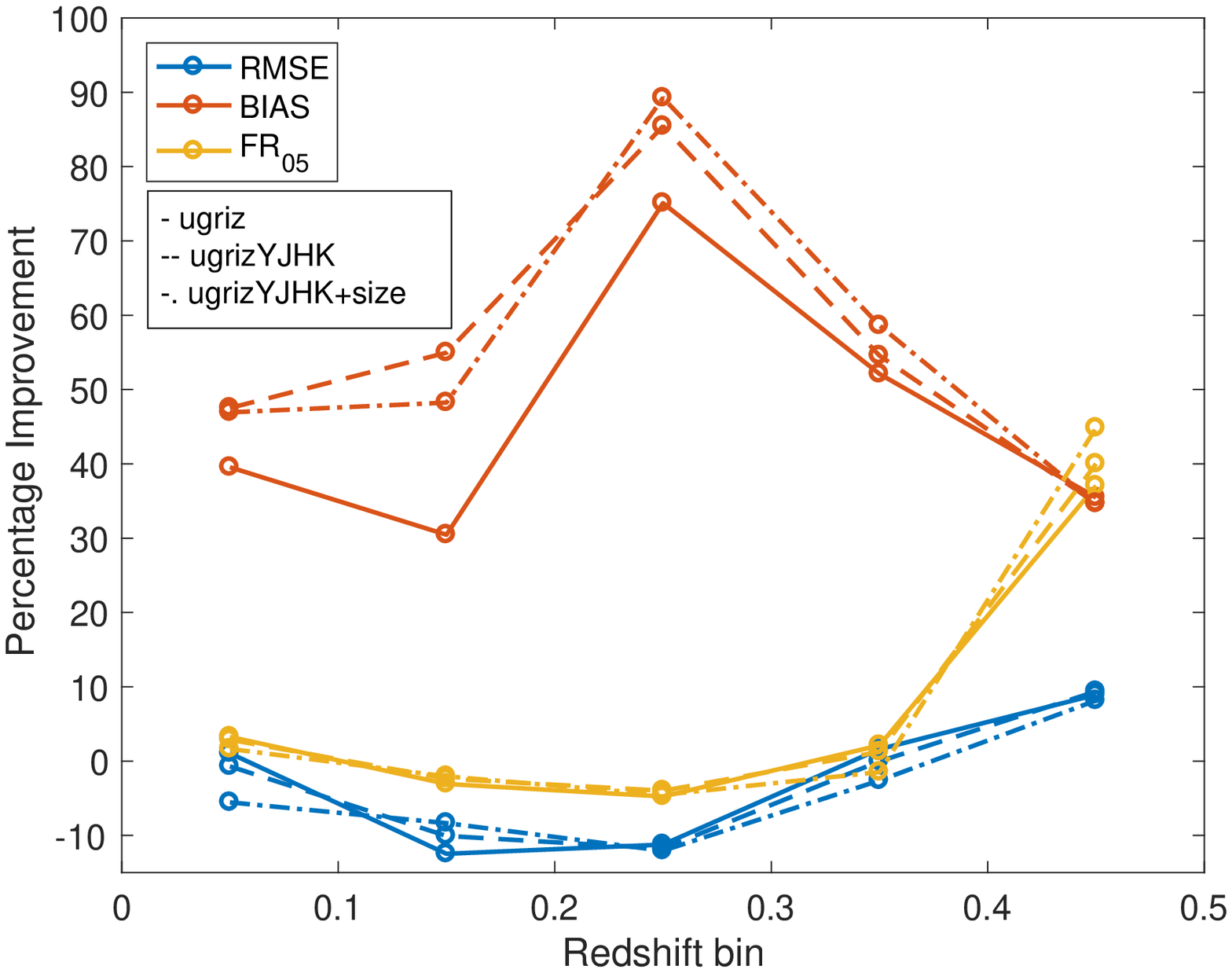}}
\newcommand{\imgzf}{\includegraphics[width=0.9\columnwidth]{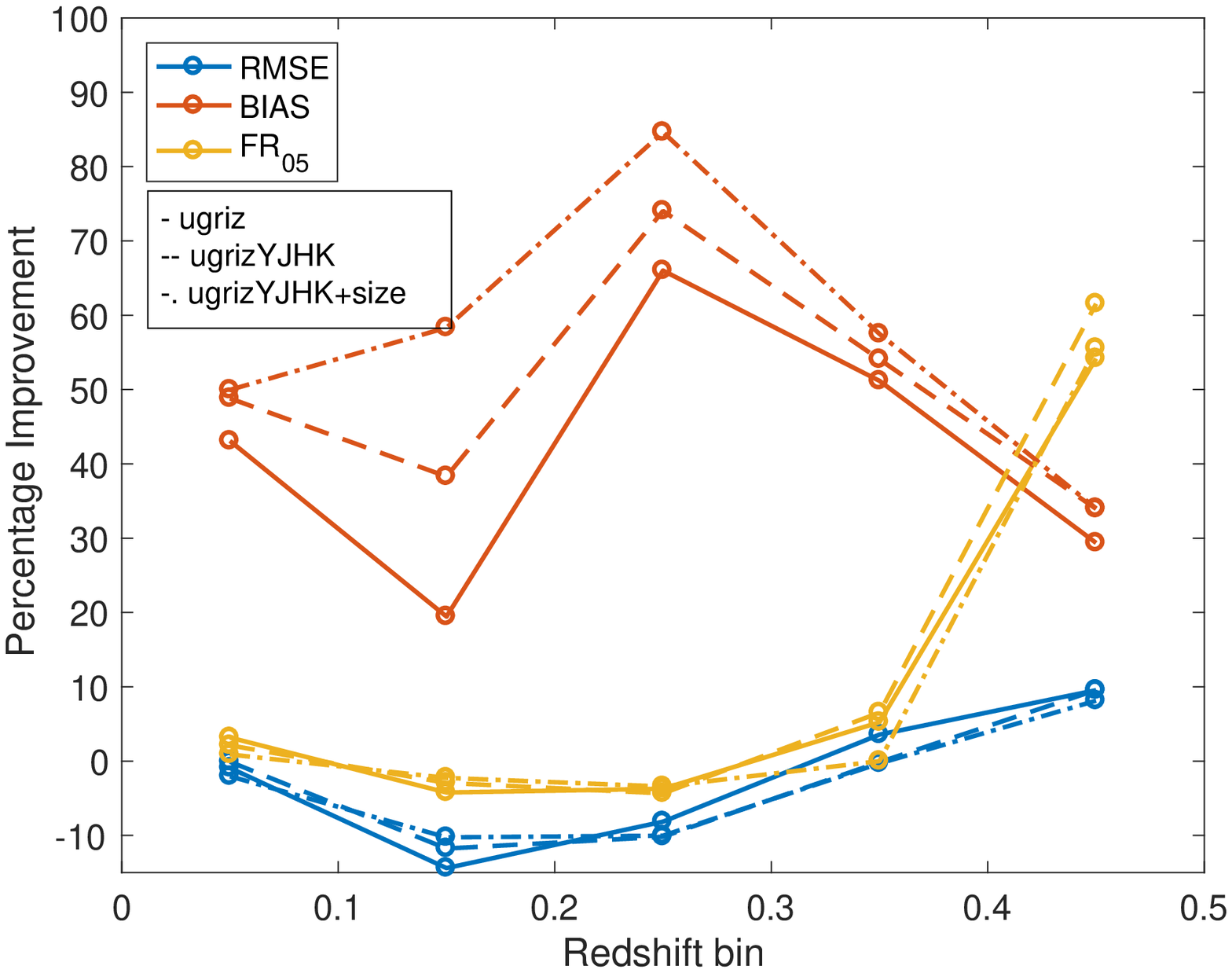}}

\begin{table*}
\centering
\begin{tabular}{cc}

  \imgze&\imgzf

\end{tabular}
\captionof{figure}{Percentage improvements of performance measures by redshift bin due to shifting the means of the photo-z PDFs compared to before the means were shifted using the same training, validation and testing objects and using the normal (left) and normalized (right) methods.The solid lines represent using the ugriz features, the dashed line represents using the ugrizYJHK features and the dash-dotted line represents using the ugrizYJHK and angular size features.}
 \label{fig:percent_improvements_shifts}
\end{table*}

Next, the percentiles were used to determine the quantiles: the quantile at a value $x$ is defined as the fraction of percentiles that are below the fraction $x$ (for example: the quantile at 0.2 is given by the fraction of objects with percentiles less than 0.2). Theoretically, for perfect sampling of a distribution we expect that for all fractions $x$ ($0<x<1$), quantile($x$) = $x$ as 20 per cent of values are expected to have percentiles less than or equal to 0.2 and so on. The calculated quantiles versus the theoretical quantiles (Q-Q) are shown in \Cref{subfig:Q-Q_before,subfig:Q-Q_after}. A Q-Q plot with a straight line indicates that the photometric redshift PDFs (and thus the means and variances) are appropriately representing the spectroscopic redshift distribution, i.e. the spectroscopic redshift values are representative of a random sampling of the photometric redshift PDFs. Deviations from this straight line indicate deviation from ideal photometric PDFs and this is quantified using the Euclidean distance ($\eta_n$). $\eta_n$ versus the shift required for the PDF (e.g. \Cref{subfig:eta-vs-Pz}) are then determined by applying multiple positive and negative shifts to the means of the PDFs, finding the respective percentiles of the spectroscopic redshifts, followed by the corresponding quantiles, and then finding $\eta_n$ of the Q-Q plots. The $p(z)$ shift that minimizes the $\eta_n$ is then taken as the shift to be applied to the photometric redshifts.

In this analysis, one half of the test data was used to produce $\eta_n$ versus $p(z)$ for each redshift bin in order to find the optimal $p(z)$ shift (\Cref{fig:badexample}). These $p(z)$ shifts were then applied to all the best fit photometric redshift values in the respective photometric redshift bins of the second half of the test data. Half of the test data was used instead of the entire set and the shifts were applied in photo-z instead of spec-z bins in order to illustrate the utility of this method when spectroscopic data is present for only a subset of the data. All spectroscopic redshifts with percentiles equal to 0 or 1 were not used for producing the Q-Q plots, as the corresponding photo-z's\textemdash defining the relevant PDFs\textemdash are considered to be outliers. The fraction of sources not within the PDF (fraction of outliers; $f_{\rm out}$) gives the number of these outliers divided by the number of objects in the relevant redshift bin. In this analysis, the $f_{\rm out}$ after applying the $p(z)$ shifts was 0 in all bins considered (0-0.5) apart from the 0.4-0.5 bin in which the $f_{\rm out}$ did not surpass $0.05$ which corresponds to $3$ objects. 

The Q-Q plots obtained after applying the photo-z shifts in all redshift bins were near to straight lines, with very low $\eta_n$values. This means that the {\sc GPz} algorithm produced appropriate variance estimates with a slight bias on the mean values. The optimal shifts found were very small in redshift bins with large numbers of data points, meaning that the mean values produced in these bins were accurate. On the other hand, the shifts were significant in the redshift bins with lower densities (see \Cref{fig:badexample}), indicating that some bias was present in these bins but that this post-processing method was able to adjust the positions of the PDFs such that they were more representative of the spectroscopic redshifts. Curves like the one shown in \Cref{subfig:Q-Q_before} which represent a lack of objects with percentiles below the given fraction for all quantiles, correspond to the PDFs generally being biased to low redshifts and thus a shift to higher redshift is suggested by the $\eta_n$ vs $p(z)$ plot. Q-Q plots that indicated that the PDFs were biased to low redshift were obtained for the higher redshift bins ($z>0.2$) while the opposite was obtained for the lower redshift bins ($z<0.1$). This can be explained by the fact that the best fit mean function will be found where the data density is highest, which is around the redshift of $z\sim 0.2$. In some redshift bins (e.g. $z>0.5$) the number of galaxies was too small for this analysis to be carried out.

 \Cref{fig:percent_improvements_shifts} gives the percentage improvements of the performance metrics by redshift bin due to shifting the PDFs (the variances are not included as they are not affected by the shifts). For all the redshift bins in which this method was applied ($z<0.5$), for all configurations of input variables and for both CSL methods we see significant improvement in the bias metric, while the other metrics only worsen slightly in some redshift bins. The bias shows the most significant improvements because this method of using the Q-Q plots to shift the PDFs specifically targets the bias. The RMSE and $\textnormal{FR}_{0.05}$ metrics both show improvements in redshift bins with lower number densities, while improvements are minimal in the highest density bins. This is because the original model was such that it fit the more dense regions better than the less dense ones, resulting in biases on both sides of the central dense region. These clear improvements demonstrate the efficiency of this post-processing method and we expect further improvements if smaller redshift bins are used.

\section{Effects of Improved Photometry}
\label{sec:impphot}

In this section we investigate and quantify the improvement in the photometric redshifts with deeper imaging data over a subset of the GAMA fields used thus far.

\newcommand{\imgs}{\includegraphics[width=0.6\columnwidth,height=125pt]{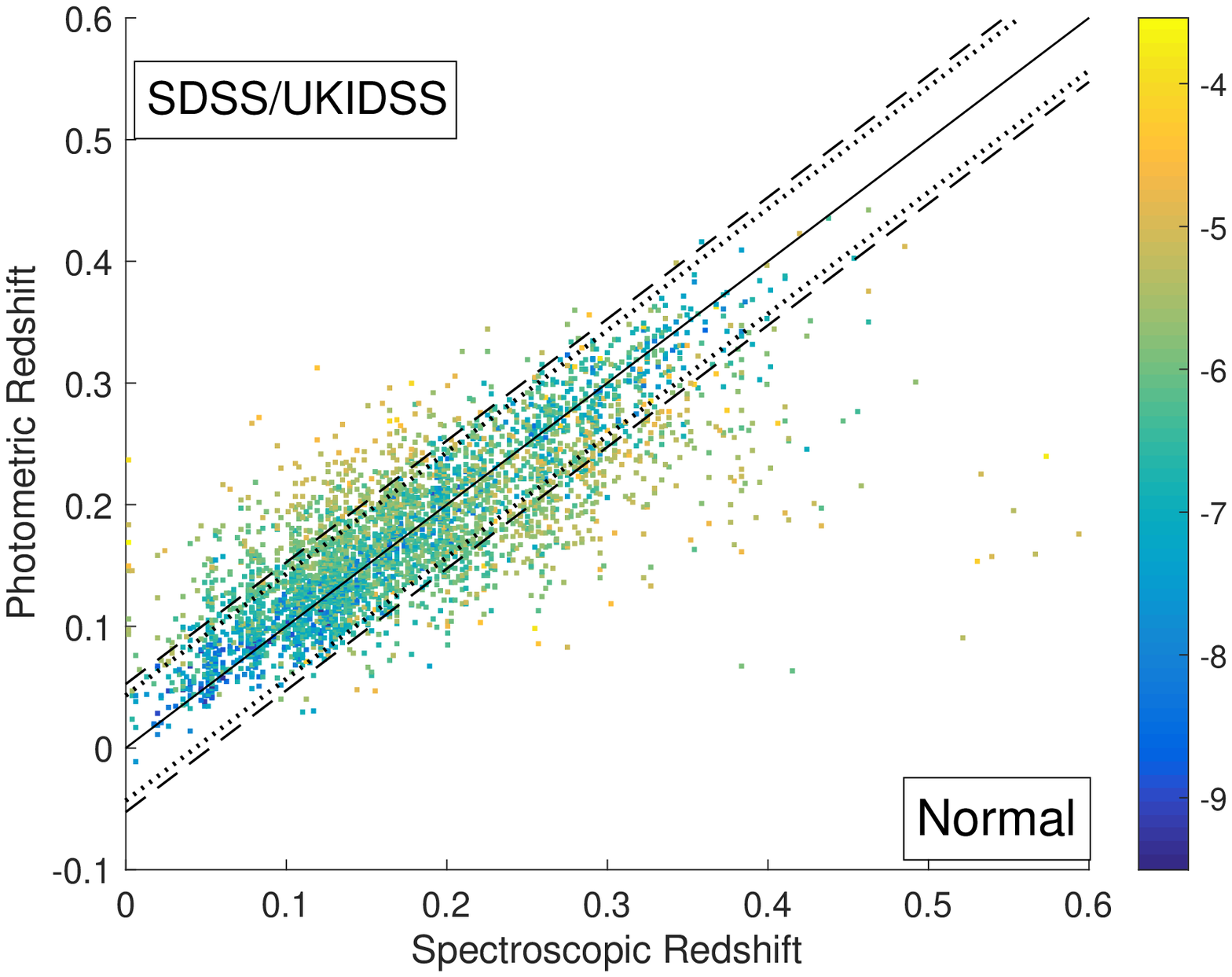}}
\newcommand{\imgt}{\includegraphics[width=0.6\columnwidth,height=125pt]{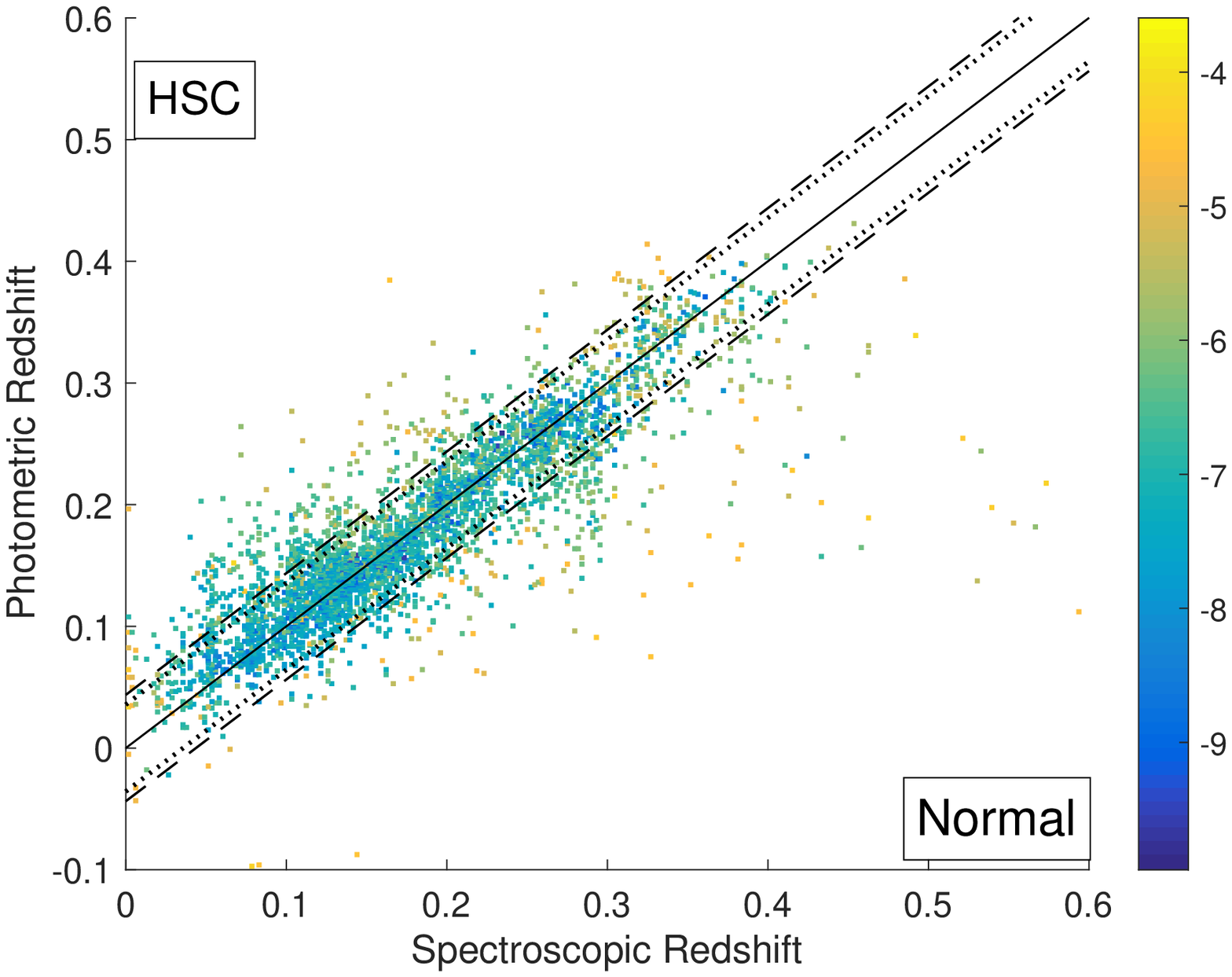}}
\newcommand{\imgu}{\includegraphics[width=0.6\columnwidth,height=125pt]{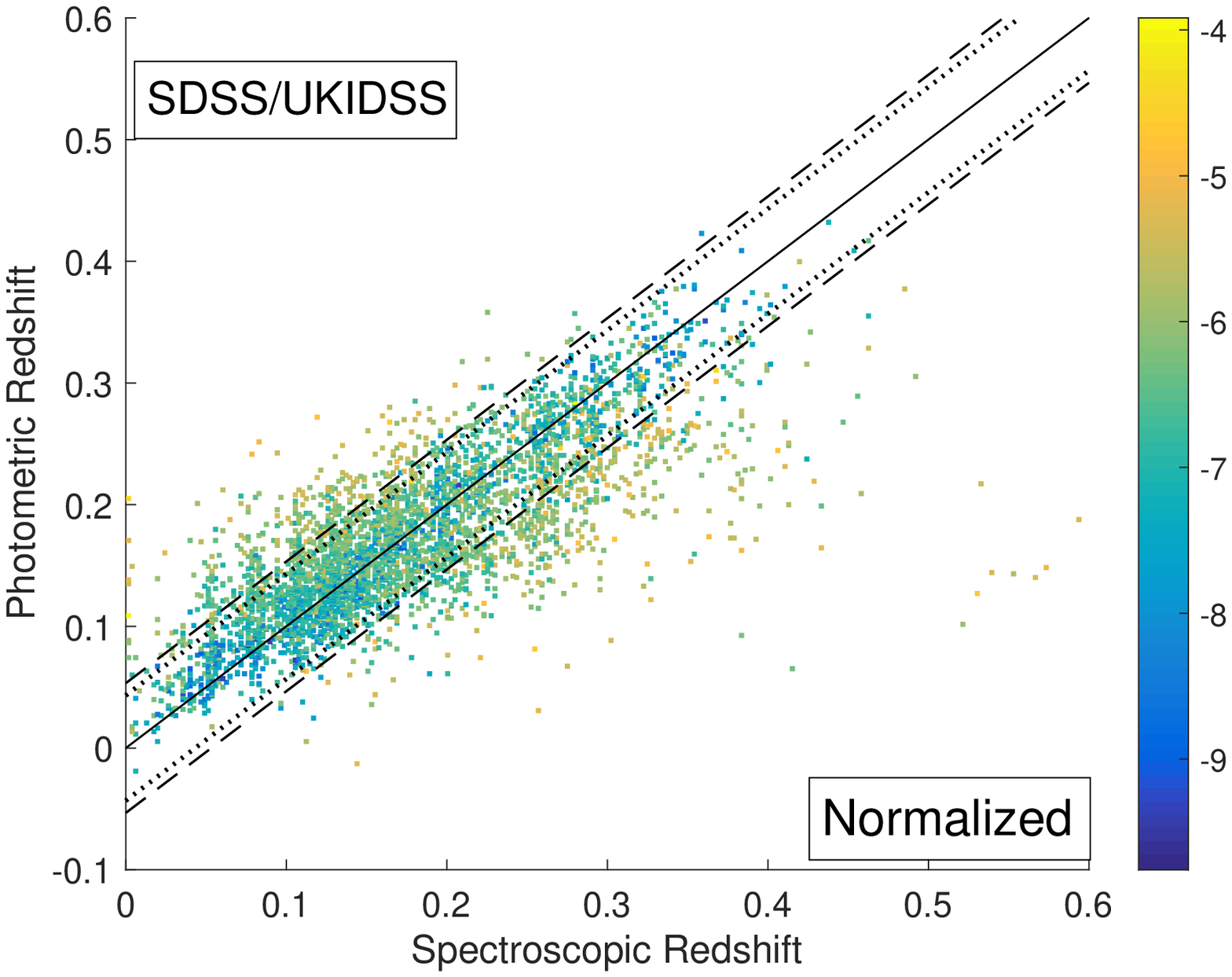}}
\newcommand{\imgv}{\includegraphics[width=0.6\columnwidth,height=125pt]{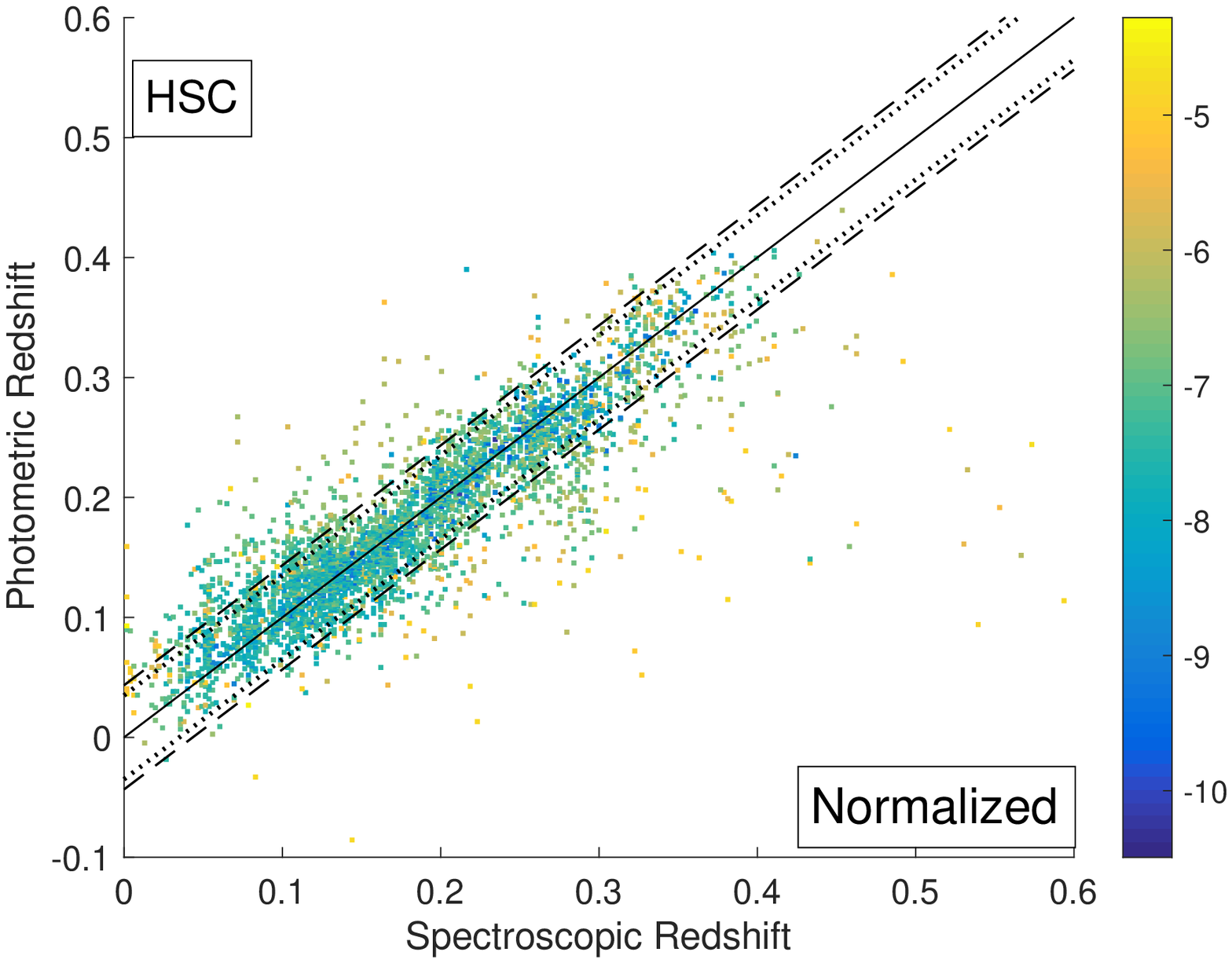}}

\begin{table*}
\centering
\begin{tabular}{cc}

  \imgs&\imgt\\
  \imgu&\imgv\\

\end{tabular}
\captionof{figure}{Photometric redshift versus spectroscopic redshift plots showing the performance of the {\sc GPz} code using SDSS/UKIDSS LAS photometry (left) and HSC photometry (right) with the CSL methods normal and normalized (in the first and second rows respectively). The colour scale represents the variance of data points in that area of the plot and the straight line is the $z=\hat z$ line.}
 \label{fig:HSCvsSDSS}
\end{table*}

\begin{table*}
\label{my-label}
\begin{tabular}{lllllllllll}
\hline
CSL Method & Survey & Features        & RMSE       & BIAS        & MLL       & $\textnormal{FR}_{\textnormal{0.15}}$ & $\textnormal{FR}_{\textnormal{0.05}}$ & Variance & Model Var & Noise Var  \\
\hline
Normal     & SDSS/      & grizY          & 0.0432 & -0.0013 & 1.70 & 99.18 & 81.25 & 0.0025 & 2.3E-05 & 0.0025 \\
           & UKIDSS LAS & ugrizYJHK      & 0.0352 & -0.0014 & 1.87 & \textbf{99.60} & 88.22 & 0.0018 & 2.1E-05 & 0.0018 \\
           &            & ugrizYJHK+size & 0.0336 & -0.0012 & 1.92 & 99.53 & 89.64 & 0.0016 & 2.1E-05 & 0.0016 \\
           & HSC        & grizy          & 0.0357 & -0.0008 & \textbf{1.94} & 99.33 & 89.14 & 0.0015 & 1.9E-05 & 0.0015 \\
\hline           
Normalized & SDSS/      & grizY          & 0.0432 & 0.0021  & 1.64 & 99.25 & 80.67 & 0.0018 & 2.3E-05 & 0.0018 \\
           & UKIDSS LAS & ugrizYJHK      & 0.0355 & 0.0010  & 1.83 & 99.58 & 88.17 & 0.0014 & 2.2E-05 & 0.0013 \\
           &            & ugrizYJHK+size & \textbf{0.0334} & \textbf{0.0007}  & 1.87 & 99.55 & \textbf{89.86} & \textbf{0.0011} & 1.9E-05 & 0.0011 \\
           & HSC        & grizY          & 0.0349 & 0.0011  & 1.85 & 99.43 & 89.64 & \textbf{0.0011} & \textbf{1.8E-05} & \textbf{0.0010}\\
           \hline
\end{tabular}
\centering
\caption{Table showing summary performance measures and variance for the HSC photometry with grizy filters and the SDSS/UKIDSS LAS photometry with all three configurations of features. The number of training, validation and testing objects are: 8047, 8048 and 4024 respectively. The best metrics and variances are highlighted. }
\label{tab:SDSSallfiltervsHSC}
\end{table*}

\newcommand{\imgzc}{\includegraphics[width=0.9\columnwidth]{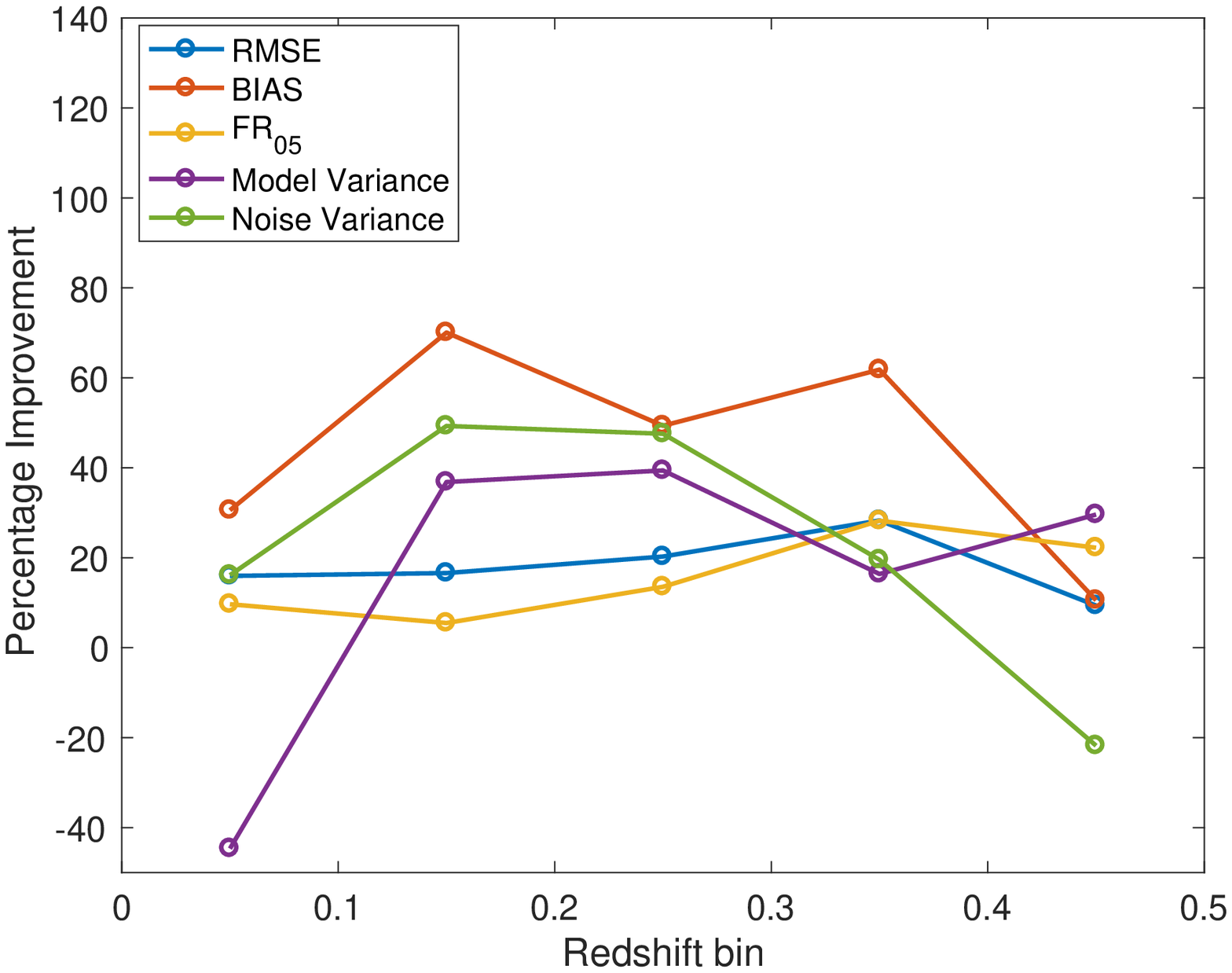}}
\newcommand{\imgzd}{\includegraphics[width=0.9\columnwidth]{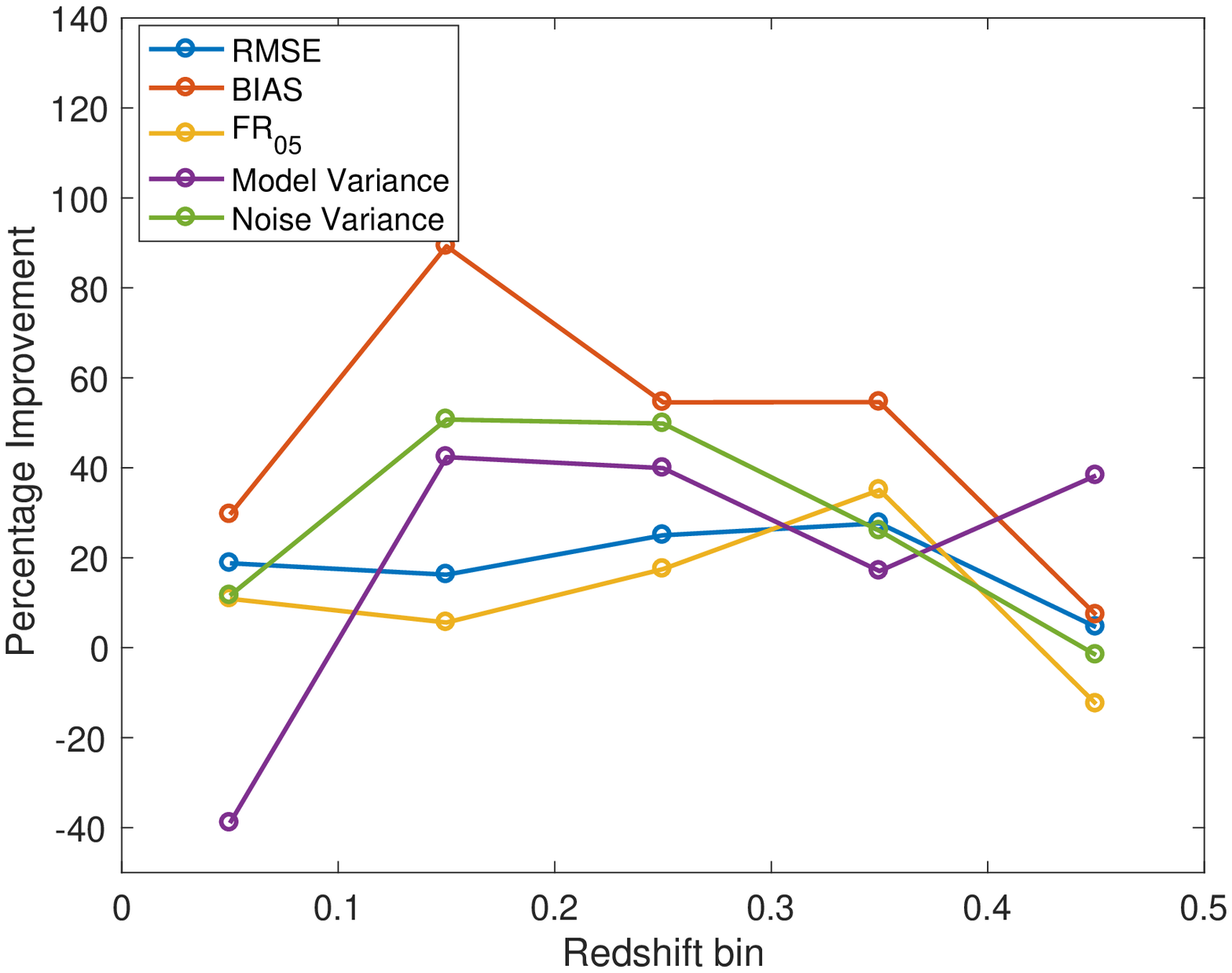}}

\begin{table*}
\centering
\begin{tabular}{cc}

  \imgzc&\imgzd

\end{tabular}
\captionof{figure}{Percentage improvements of performance measures and variances by redshift bin due to the use of HSC grizy filters compared to SDSS/UKIDSS LAS grizY filters using the same training, validation and testing objects and using the normal (left) and normalized (right) methods.}
 \label{fig:percent_improvements_HSC}
\end{table*}

The Hyper Suprime-Cam Subaru Strategic Program (HSC-SSP) Data Release 1(\citealt{aihara2017first}) provides photometry in grizy filters with magnitude errors that are an order of magnitude smaller than the SDSS/UKIDSS LAS photometry. We crossmatched the positions of the galaxies used for the previous analyses with the HSC galaxies and combined the corresponding photometry with the GAMA spectroscopy. After eliminating spec-z's with NQ<3 and removing any objects with missing SDSS/UKIDSS LAS or HSC photometry we found that only 20253 GAMA objects were matched to HSC objects. This was because only limited portions of the GAMA fields were covered in the HSC-SSP survey (see \citealt{aihara2017first}). 

The {\sc GPz} algorithm with 100 basis functions and modelling of heteroscedastic noise was then used to estimate the photometric redshifts from both the HSC grizy photometry and the SDSS/UKIDSS LAS grizY photometry for an identical set of galaxies. It should be noted that the Y filter from the UKIDSS LAS photometry is of a different shape to the y filter from the HSC photometry, but they are similar enough to be used for comparison. Photometric redshift versus spectroscopic redshift plots are shown in \Cref{fig:HSCvsSDSS}, the percentage improvements of the performance measures and variances by redshift bin are shown in \Cref{fig:percent_improvements_HSC}, and metrics are given in \Cref{tab:SDSSallfiltervsHSC}. It is evident from these figures that the HSC photometry produces a tighter distribution than the SDSS/UKIDSS LAS photometry, and the metrics showing significant improvements when the HSC data is used. In addition to improved metrics, the model and noise variances are also improved. As previously discussed, improved precision of the input data should decrease the noise variance as the spread of the data should decrease. The model variance also improves as the algorithm is much more confident about the model fit with the more precise data.

Next, all available filters as well as size data were added to the estimates obtained using the {\sc GPz} algorithm from the GAMA SDSS/UKIDSS LAS dataset. The results are summarized in \Cref{tab:SDSSallfiltervsHSC}. We see that although the HSC data clearly outperforms the SDSS/UKIDSS LAS data when only five filters are used, it does very similarly to the SDSS/UKIDSS LAS data when near-IR data are added and is outperformed in most metrics by a small margin when size data is also added. 

Considering the limited size of the data set, the improvements in the metrics provided by the improved quality of the photometry are very significant and thus, further improvements in photometry over large survey regions\textemdash as will be provided by future surveys such as the Large Synoptic Survey Telescope (LSST; \citealt{lsstbook}) and {\em Euclid} (\citealt{Laureijs2011euclid})\textemdash will have significant impacts on the ability of the {\sc GPz} algorithm to accurately predict the photometric redshifts of galaxies.  

\section{Conclusions}\label{sec:conclusions}

In this paper, methods of obtaining improved photometric redshift estimations from the {\sc GPz} machine learning algorithm were investigated. These methods were introducing near-IR magnitudes and angular size features, post-processing the results by shifting the photo-z estimates based on their Q-Q plots and utilizing photometry with higher precision. It was found that the inclusion of near-IR (YJHK) filters and angular size data in the training, validation and testing of photometric redshift estimation resulted in significantly improved accuracy, and thus, when available, this data should be utilized. The process of shifting the probability distributions of the estimated redshifts by minimizing the $\eta_n$ value has proven to substantially improve the bias of the estimated photometric redshifts. Therefore, when a suitable spectroscopic sample is available, this method could be applied to supply additional accuracy to the predictions from {\sc GPz} and other methods.  Finally, we see that improvements in the accuracy of the photometry improved the accuracy of the photometric redshifts, to a very similar extent as adding the near-IR and angular size data, and therefore, work should continue to be done to improve the quality of the photometric data obtained.

It is worth mentioning that we have targeted galaxies predominantly at $z<0.5$ in this study, where one might expect the size information to have more of an influence, but where one might also expect the near-infrared filters to add a comparatively smaller amount of information compared to $z> 1$, where the 4000\AA\space break moves out of the visible wavelength filters. In a future paper we will explore this issue by combining the deeper visible wavelength data \citep[e.g.][]{aihara2017first} with deeper near-infrared data  over the well studied extragalactic deep fields.

\section*{Acknowledgements}

ZG is supported by a Rhodes Scholarship granted by the Rhodes Trust. IAA would like to acknowledge the support of King Abdulaziz City for Science and Technology.   

GAMA is a joint European-Australasian project based around a spectroscopic campaign using the Anglo-Australian Telescope. The GAMA input catalogue is based on data taken from the Sloan Digital Sky Survey and the UKIRT Infrared Deep Sky Survey. Complementary imaging of the GAMA regions is being obtained by a number of independent survey programmes including GALEX MIS, VST KiDS, VISTA VIKING, WISE, Herschel-ATLAS, GMRT and ASKAP providing UV to radio coverage. GAMA is funded by the STFC (UK), the ARC (Australia), the AAO, and the participating institutions. The GAMA website is http://www.gama-survey.org/. 

The Hyper Suprime-Cam (HSC) collaboration includes the astronomical communities of Japan and Taiwan, and Princeton University. The HSC instrumentation and software were developed by the National Astronomical Observatory of Japan (NAOJ), the Kavli Institute for the Physics and Mathematics of the Universe (Kavli IPMU), the University of Tokyo, the High Energy Accelerator Research Organization (KEK), the Academia Sinica Institute for Astronomy and Astrophysics in Taiwan (ASIAA), and Princeton University. Funding was contributed by the FIRST program from Japanese Cabinet Office, the Ministry of Education, Culture, Sports, Science and Technology (MEXT), the Japan Society for the Promotion of Science (JSPS), Japan Science and Technology Agency (JST), the Toray Science Foundation, NAOJ, Kavli IPMU, KEK, ASIAA, and Princeton University. 

This paper makes use of software developed for the Large Synoptic Survey Telescope. We thank the LSST Project for making their code available as free software at http://dm.lsstcorp.org. 

The Pan-STARRS1 Surveys (PS1) have been made possible through contributions of the Institute for Astronomy, the University of Hawaii, the Pan-STARRS Project Office, the Max-Planck Society and its participating institutes, the Max Planck Institute for Astronomy, Heidelberg and the Max Planck Institute for Extraterrestrial Physics, Garching, The Johns Hopkins University, Durham University, the University of Edinburgh, Queens University Belfast, the Harvard-Smithsonian Center for Astrophysics, the Las Cumbres Observatory Global Telescope Network Incorporated, the National Central University of Taiwan, the Space Telescope Science Institute, the National Aeronautics and Space Administration under Grant No. NNX08AR22G issued through the Planetary Science Division of the NASA Science Mission Directorate, the National Science Foundation under Grant No. AST-1238877, the University of Maryland, and Eotvos Lorand University (ELTE) and the Los Alamos National Laboratory.

Based [in part] on data collected at the Subaru Telescope and retrieved from the HSC data archive system, which is operated by Subaru Telescope and Astronomy Data Center, National Astronomical Observatory of Japan.

\bibliographystyle{mnras}
\bibliography{refs}

\bsp	
\label{lastpage}
\end{document}